# Hierarchical Mobility Label Based Network: System Model and Performance Analysis

Oleg Berzin*

*Verizon Communications, Philadelphia, PA, USA, E-Mail: oleg.berzin@verizonwireless.com

*Abstract* – **Hierarchical Mobility Label Based Network (H-MLBN) is a new approach to the network layer mobility management problem that relies on MPLS-aware control plane and MPLS-based forwarding plane to provide IP mobility support for IPv4 and IPv6 mobile hosts and routers while being able to ensure optimal traffic delivery between the communicating devices. The hierarchical system is capable of both macro- and micro-mobility support without the use of Mobile IP and its derivatives thus eliminating the user and network facing performance penalties associated with triangular routing and bi-directional tunneling. This paper presents a system model and provides performance analysis for H-MLBN and compares its performance with the Mobile IP based schemes. The results indicate significant performance improvements in the forwarding plane traffic delivery as well as the control plane network update costs.**

## I. Introduction

The Hierarchical Mobility Label Based Network (H-MLBN) integrates the layer 3 mobility control plane and the MPLS [10] forwarding plane in order to achieve optimal traffic delivery between the communicating mobile devices. This is achieved by using Mobility Labels (stacked MPLS labels) to represent the current location of a mobile device in the network. Mobility Labels are associated with the IP addresses of mobile hosts or IP prefixes served by mobile routers to form Mobility Bindings. Mobility Bindings in turn are distributed using MPLS-aware H-MLBN control plane protocol (based on MP-BGP [8], [5]) to other H-MLBN network nodes to identify the current location of the mobile device or the next Label Switched Path (LSP) segment on the path to the mobile device. The traffic delivery is then based on MPLS label stack and follows the optimal network path. The concept of the Mobility Label Network (MLBN) was introduced in [1] and the H-MLBN is described in detail in [2].

H-MLBN involves a hierarchical network architecture for both the control and forwarding planes as well as the regionalized network structure. The network coverage area is divided into Mobility Regions which are grouped into larger Mobility Areas. The following definitions of the architectural entities are used in H-MLBN:

*Label Edge Router (LER)* – an edge node in MLBN. LER connects to RAN using L2 grooming network. Each RAN may be terminated at a logical L3 interface of the LER which represents an IP sub-net. LER implements a Mobility Support Function (MSF) and peers with Area MRR using MP-BGP.

*Mobility Support Function (MSF)* – a set of processes executing at LER and responsible for mobile device (host or router) registration, Mobility Label assignment, Mobility Binding creation and distribution using a network update.

*Mobility Region* – a collection of the RAN cells or clusters served by a single MSF residing in the MPLS LER node.

*Mobility Area* - a collection of Mobility Regions aggregated by the Area LER (ALER).

*Area LER (ALER)* – MPLS aggregation node that implements MSF and participates in the packet forwarding.

*Area Mobility Route Reflector (AMRR)* – a Mobility Route Reflector serving the Mobility Area. AMRR does not participate in the packet forwarding and performs control plane (signaling) functions using MP-BGP.

*Mobility Label (ML)* – a MPLS label that is associated with a mobile prefix (IPv4 or IPv6 MN's address or network prefix served by a mobile router). Mobility label is used to represent current network location of a mobile device. Mobility Labels are assigned by LERs and used as inner labels in the MPLS label stack.

*Mobility Binding (MB)* – an association between the mobile prefix, and the Mobility Label. Mobility Bindings are distributed using MP-BGP and encoded using a specific Network Layer Reachability Information (NLRI) format [7].

*Network Update* – a process of distributing Mobility Bindings throughout MLBN.

Fig. 1 illustrates a Hierarchical MLBN (H-MLBN) with nine Mobility Regions (MR) served by the LER nodes at the edge level of the MPLS network. Each LER connects to a set of RAN cells or clusters using the layer 2 access/grooming network such as Ethernet or ATM (not shown) and executes a Mobility Support Function (MSF). Each RAN cell or cell cluster is terminated at a layer 3 logical interface of the LER controlled by the MSF.

At the aggregation level the edge level LER nodes are connected to the Area LER nodes (ALER) thus forming the Mobility Areas. Fig. 1 shows three Mobility Areas each comprising of three Mobility Regions.

At the control plane level Fig. 1 shows three Area Mobility Route Reflectors (AMRR) serving their respective Mobility Areas and peering directly with the corresponding MSF LER and the ALER nodes within the area. The AMRR nodes form a H-MLBN control plane. The ALER and AMRR nodes are connected to the MPLS backbone network that consists of the LSR nodes.

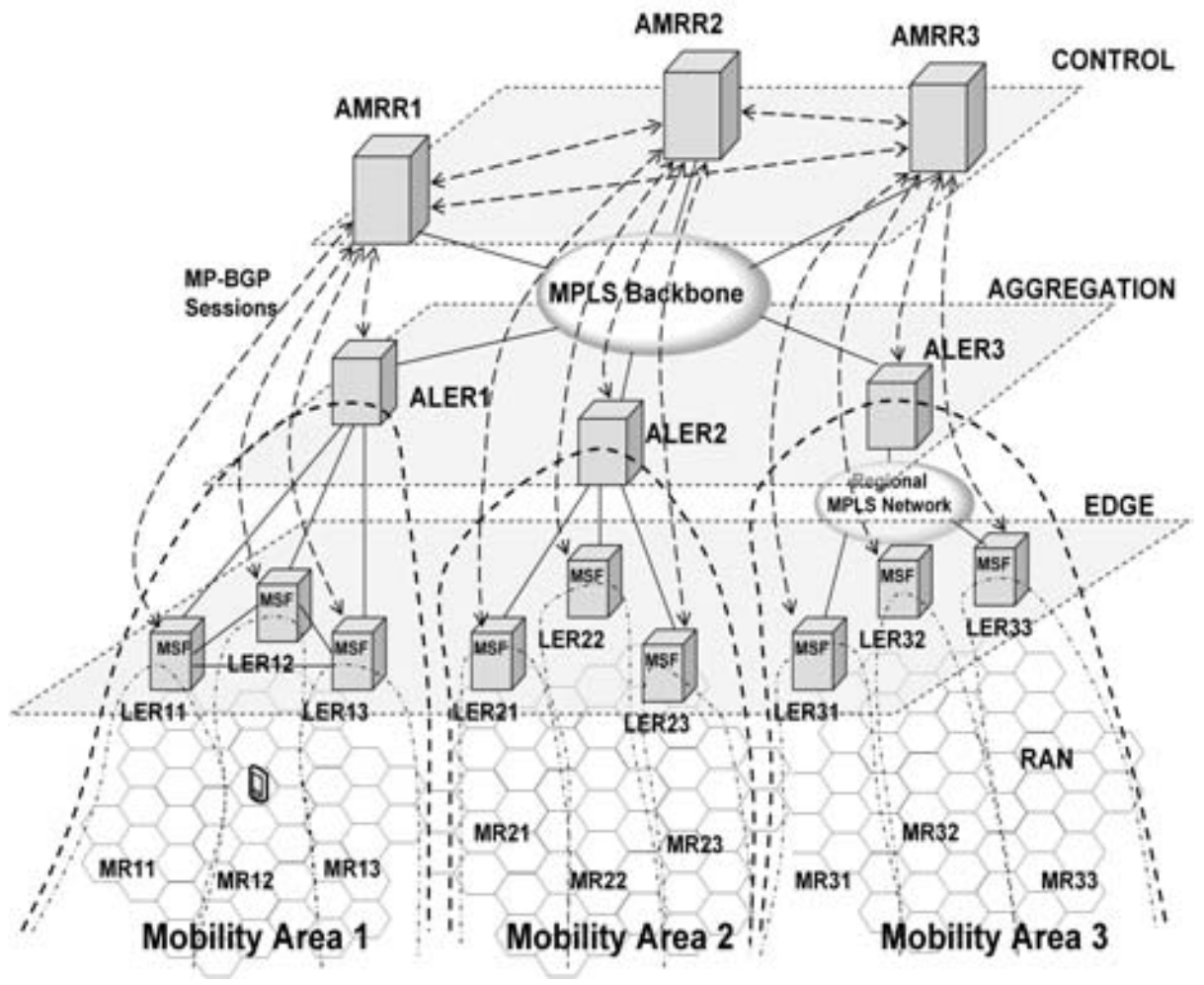

Figure 1. Hierarchical Mobility Label Based Network.

The mobile hosts and routers register only with the serving MSFs at the edge level of the H-MLBN. The MSF Discovery and Registration protocols defined in [3] do not extend beyond the MSF LER nodes. The registration with a MSF results in the assignment of the Mobility Label and generation of the corresponding Mobility Binding as described in [3].

Upon the completion of the registration the LER nodes at the edge level update the AMRR using a corresponding network update [2-3] carrying the Mobility Binding information for the registered mobile devices.

This paper presents a system model and provides performance analysis for H-MLBN and compares its performance with the Mobile IP based schemes. The rest of the paper is organized as follows: Section II describes related work, Section III develops the movement model for mobile devices, Section IV develops traffic models for the forwarding and control planes, Section V develops performance metrics and presents comparative analysis, Section VI provides numerical and performance results and Section VII offers the conclusions.

## II. Related Work

### A. Mobile IP Macro-mobility

*Mobile IPv4*: MIPv4 [4] provides macro mobility management for mobile hosts using IPv4. The main entities in MIPv4 are the Mobile Node (MN), the Correspondent Node (CN), the Home Agent (HA) – the router that owns the original home sub-net, the Foreign Agent (FA) – the router that owns the sub-net to which the MN has moved, and the Care-of-Address (CoA) – the IP address that belongs to the FA and that is used to represent the MN while it is located in the foreign sub-net. The basic operation of MIPv4 requires traffic tunneling between FA and HA (IP-IP or GRE [6]) and results in triangular routing when traffic from CN to MN needs to first go through HA and then use the tunnel from HA to FA to reach the MN. From MN to CN the traffic may follow directly. However, due to the requirement that the source IP address of the packets sent by MN is its Home Address (which is topologically incorrect in the foreign sub-net) the direct MN to CN path may be impossible due to the ingress packet filtering [12]. To overcome this reverse tunneling is used – MN sends packets to FA, FA tunnels packets to HA and HA sends data to CN. Thus with reverse tunneling basic MIPv4 suffers from bi-directional suboptimal routing.

*Mobile IPv4 with Route Optimization*: ROMIPv4 [18] allows CN to send packets directly to MN without going through HA-FA tunnel thus eliminating triangular routing. ROMIPv4 however imposes significant requirements on CN (which is any IPv4 host on the Internet) that make it an unlikely deployment choice for practical applications. Specifically, CN is required to support MIPv4 binding processing as well as to use tunneling to communicate with MN. In addition, ROMIPv4 requires that MN registers and authenticates with CN (just as it does with HA) and thus poses a problem of distribution and management of relevant security information to every CN MN communicates with. ROMIPv4 also imposes additional complexity and processing load on HA that is now required to keep track of CNs and update them with new binding information as MNs move about.

*Mobile IPv4 Network Mobility:* NEMOv4 [19] is a set of extensions to MIPv4 that allow a mobile router (MoR) to register its LAN subnets during MIPv4 registration process. NEMOv4 may operate through a MIPv4 FA or in a Collocated Care-of-Address (CCOA) mode. Using the CCOA mode MoR registers its CCOA, its Home Address and the IP prefixes directly with the MIPv4 HA. MoR then establishes a direct MoR-HA tunnel and the HA is responsible for forwarding the traffic destined to the IP devices attached to MoR through the tunnel identified by the registered CCOA. NEMOv4 does not support route optimization and is subject to triangular routing or bi-directional tunneling resulting in suboptimal traffic routing. In addition NEMOv4 imposes increased load on HA which has to maintain direct MoR-HA tunnels (as opposed to the FA-HA tunnels that may carry multiple MIPv4 sessions).

*Mobile IPv6:* MIPv6 [11] provides macro-mobility support for IPv6. It improves MIPv4 by eliminating the need for FA and use of IPv6 Link Local (LLOC) address instead of CoA. MIPv6 allows MN to use its CoA as a source IP address in all packets it sends thus overcoming the ingress packet filtering issue. MIPv6 provides direct support for route optimization by allowing MN to register itself with CN and update it with its mobility bindings. However just like in ROMIPv4 route optimization for MIPv6 requires that CN (any node on the Internet) support MIPv6 and special IPv6 extensions such as routing header and destination option. In addition, route optimization requires a separate return routability procedure executed on both MN and CN and partly via HA in order to ensure that the packets are sent to a correct MN by CN. These additional requirements on CN are fairly significant and may be considered as obstacles in implementing MIPv6 with route optimization. Another mode of operation for MIPv6 is *bi-directional tunneling* in which CN does not need to support MIPv6 and all traffic is tunneled through HA in both directions via suboptimal routing path. In addition, elimination of FA results in increased load on HA that now needs to manage individual security associations and tunnels for every MN.

*Mobile IPv6 Network Mobility:* NEMOv6 [20] is part of MIPv6 and enables a stationary or a mobile router to register with HA, receive a home address and register the network prefixes it serves with HA. The mobile router establishes a bi-directional tunnel with it's HA. The HA binds the network prefixes it receives from the mobile router to the router's care-of-address reachable via a tunnel, and advertises the prefixes into the serving IP network. When packets are sent to devices connected to a mobile router and residing in the registered prefixes the IP network routes the packets to the HA and the HA tunnels the packets to the mobile router which forwards them to the destination. Just like in NEMOv4 traffic for the destinations served by a mobile router has to use a suboptimal path via HA.

*B. Mobile IP Micro-mobility*

*Mobile IPv4 Regional Registration:* RRMIPv4 [21] aims to minimize the frequency of HA re-registrations due to MN movements. It proposes a hierarchical FA structure in which Regional FA (RFA) and Gateway FA (GFA) act as proxy HA systems by relaying their addresses as CoA addresses for MN. MN registers with RFA and while it changes its location within the serving area of that RFA no HA re-registration is required. While RRMIPv4 provides micro-mobility in a sense that it hides MN movements within a geographical region from HA it does not address optimal routing. As a matter of fact it introduces its own suboptimal routing structure rooted at a given RFA or GFA as the traffic must visit these nodes to be tunneled to MN. RRMIPv4 is still subject to triangular routing, ingress packet filtering and bi-directional tunneling.

*Hierarchical Mobile IPv6:* HMIPv6 [22] uses Mobility Anchor Point (MAP) as a local HA in order to minimize the frequency of MN re-registrations to the real HA thus hiding the MN movements within the service area of a given MAP from HA. MN obtains two CoA addresses LCOA as in MIPv6 and RCOA (Regional COA – belongs to a MAP). MN registers its LCOA with a MAP and RCOA with its HA. MAP is responsible for tunneling packets destined to the MN's Home Address to its LCOA. As in RRMIPv4 HMIPv6 has its own suboptimal routing structure rooted at a given MAP and does not provide optimal traffic routing (at least in the bi-directional tunneling mode). MIPv6 route optimization applied in the HMIPv6 environment still requires MIPv6 and special IPv6 header/options support on a CN.

*C. Network-based Mobile IP*

*Proxy Mobile IPv6* – PMIPv6 [26] eliminates the need for MIP support on the MN and shifts all mobility control functions to the network. It uses a Mobile Access Gateway (MAG) to execute MIPv6 signaling on behalf of the MN with a Local Mobility Anchor (LMA). MAG is roughly equivalent to a MIP FA and LMA to a MIP HA. When MN enters a PMIPv6 domain it attaches to a MAG which authorizes MN's access and then registers MN with the LMA. The LMA assigns the home network prefixes for the MN and returns this information in the MIPv6 signaling to the MAG. MAG configures the attached interface of the MN with the home network prefixes received from the LMA. When traffic is sent to one or more of the home network prefixes on the MN, LMA tunnels the traffic to the serving MAG and MAG forwards it to the MN. When MN changes its location, the previous MAG is responsible for detecting the change and signaling MN's detachment to the LMA. When MN attaches to the new MAG, the MAG updates the binding information for the MN on the LMA and continues to advertise the same local link level addressing and the home network prefixes as the previous MN. PMIPv6 does not support route optimization but does support a limited form of optimal routing for a case when both the CN and the MN are attached to the same MAG. Therefore in general, PMIPv6 has to use bi-directional tunneling that results in sub-optimal routing as in other MIP based schemes. PMIPv6 can support IPv6, IPv4 and dual-stack MN's by tunneling IPv4 binding information within PMIPv6.

*D. Role of MPLS in Support of Micro-Mobility*

The use of MPLS to provide micro-mobility support is attractive as MPLS traffic forwarding is not based on the IP addresses but on the MPLS labels instead. However, allocation of the MPLS labels inside the network nodes (routers) following the movements of mobile nodes is a process that represent additional and in some cases significant overhead on the control plane. The label distribution process results in the construction of the Label Switched Paths (LSPs) that connect the endpoints reachable via the LSP.

We distinguish two classes of label distribution processes: flow-driven and topologically-driven. In the flow-driven approach the labels are allocated on-demand for a communication path between a given source and a given destination, and the label allocation process is based on a connection-oriented signaling protocol such as ReSource reservation Protocol (RSVP) [29]. In the topologically-driven approach the distribution of labels is a function of the routing protocol used in the network. Once the routing protocol converges the labels are assigned to every possible reachable IP source-destination pair that is part of the network topology. Thus a full logical mesh of LSPs is created almost immediately after the logical network topology is established and the LSPs between the sources and destinations are preconfigured by a protocol such as the Label Distribution Protocol (LDP) [30].

In application to providing mobility management using MPLS the differences resulting from the two approaches may be dramatic. Specifically the flow-driven approach may result in a significant signaling overhead at the control plane as a large number of frequently moving mobile nodes would require a high frequency of LSP set-ups or tear-downs following the movements of the mobile nodes.

*E. MPLS Micro-mobility*

The MPLS Micro-Mobility [12]-[17], [23] combines Mobile IP architecture and MPLS traffic forwarding to implement mobility support solutions in which Mobile IP is used for macro mobility and MPLS is used to support micro-mobility in the part of the network that interfaces with mobile hosts – MPLS domain. Micro-mobility reflects mobile host movements that can be handled without the re-registration with the Mobile IP HA.

In LEMA MPLS-based Micro-Mobility (LMM-MPLS) [23] the mobile host registers with a hierarchical set of special MPLS Label Edge Routers referred to as Label Edge Mobility Agents (LEMA). The LEMA at the top of the hierarchical set is registered with the Mobile IP HA as the FA for the MN. A mobile host receives advertisements from the Access Routers (AR) containing the addresses of a subset of LEMAs and their relationship in the LEMA hierarchy. Mobile hosts chooses a set of LEMAs to register with and the LEMA at the top of the registration tree registers the mobile host with the HA. HA tunnels all packets from CN to MN to the top level LEMA as in regular Mobile IP. Once packets are received and de-encapsulated from the tunnel at the LEMA, the packets are sent on the MPLS LSP to the network location of the MN using the MPLS labels assigned to the MN's IP address as the result of the registration process. As the MN moves to new locations, the hand-off procedures are invoked that start with the MN requesting the hand-off and the LEMA(s) performing the set of signaling steps resulting in the redirection of the MPLS LSP from the old serving LEMA to the new serving LEMA. If the MN movement results in a condition in which the old top level LEMA can no longer serve MN, MN re-registers with the new hierarchical set of LEMA(s) and the top level LEMA is registered as the FA with the Mobile IP HA. Although in [23] MPLS Micro-Mobility makes use of the MPLS traffic forwarding it still is an extension of Mobile IP and requires mobile hosts to implement a complex logic involving a set of registrations such as a registration to the local serving Access Router, registrations to the hierarchy of LEMA(s) and the Mobile IP registration to the HA. The scheme in [23] requires heavy use of signaling starting from the need at the MN to understand the LEMA serving network hierarchy and maintain the registration chain of the serving set of LEMA(s), the requirements for the LSP redirection during the hand-off and finally a requirement for every LEMA node in the communications path to participate in the mobility signaling. The scheme in [23] does not address optimal traffic routing as the overlay LEMA network introduces its own suboptimal tree rooted at the lowest LEMA in the hierarchy with respect to a given MN. In order to provide for truly optimal traffic routing within the MPLS domain every node in the domain would have to be a LEMA (including the AR). In addition, it does not offer support for mobile routers.

In Micro Mobile MPLS (MM-MPLS) [16], [17] a two-level hierarchy in the MPLS domain is proposed where a LER/FA (Label Edge Router/Foreign Agent) is placed at the edge of the network and the LERG (Label Edge Router Gateway) is connecting the MPLS domain to the Core IP network that contains Mobile IP HA. LERG acts as a proxy HA for MN and registers its own address with HA on behalf of MN. The rest of the MPLS nodes (LSR – Label Switch Routers that are other than the LER/FA and LERG) in the domain do not need to understand Mobile IP. However these nodes do need to participate in mobility management on a per-mobile device basis since all MN movements within the MPLS domain result in the establishment of new or redirection of existing LSPs between the serving LER/FA and LERG by means of the RSVP signaling protocol (explicit LSP setup), thus affecting the scalability of the proposal. The Fast Handoff mode requires a heavy use of signaling to setup new LSPs from LERG to the new anticipated LER/FA nodes neighboring the current serving LER/FA for MN which further impacts scalability. In the Forwarding Chain mode a set of LER/FAs is advertised to MN and MN is required to choose the LER/FAs from the set it has to register with. In addition, following the movements of MN a series of LSP redirections (from old LER/FA to new LER/FA) by means of explicit LSP setup using RSVP is executed to support traffic continuity during handoffs. As in all other micro-mobility proposals [15] and [17] do not directly address optimal traffic delivery as it introduces its own suboptimal routing structure rooted at a given LERG especially for MN-MN communications.

In [13] a concept of a Micro-cell Mobile MPLS (MCM-MPLS) is introduced in order to address the issues of suboptimal traffic routing within the MPLS domain. MM-MPLS is an evolution of the Hierarchical Mobile MPLS (as in [15], [17]) in which intermediate LSRs between the LER/FA and LERG (or MPLS-aware FDA) are required to keep track of the MN home addresses and the associated MPLS labels (distributed via RSVP). This allows to introduce a concept of a crossover LSR to handle the handoffs from one micro-mobile domain to another without having to redirect LSPs from old LER/FA to new LER/FA. When MN registers with new LER/FA a RSVP signaling message is sent to LERG. At that time an intermediate LSR that has the information about the old LSP for the same MN home address intercepts the RSVP message and redirects the old LSP to the new LER/FA thus avoiding the use of the old LSP and the associated suboptimal path. This however, comes at the expense of significant added complexity and processing in the MPLS domain where every LSR is required to maintain the state information for every MN and explicit LSP signaling on a per-MN basis is required to handle both the initial MN registrations and the subsequent handoffs.

In [14] (RAN-MM-MPLS) the concept of MPLS micro-mobility domain is expanded to include the Radio Access Network (RAN) by requiring RAN Base Stations (BS) to act as IP/MPLS nodes (LER) as well as Mobile IP FA. A Gateway (GW) is used to connect the MPLS/RAN to other IP networks. This gateway acts as both MPLS LER and Mobile IP HA. The authors in [14] recognize scalability issues related to the per-MN LSP management via RSVP and propose the use of LDP (Label Distribution Protocol) to establish any-to-any pre-constructed LSP connectivity within the MPLS domain. Since the BS nodes and the GW are all MPLS-aware the Mobile IP signaling is integrated with MPLS and is used to map the mobile nodes' home addresses to existing LSPs. This proposal, however, does not address optimal traffic delivery (specifically for MN-to-MN communications) as all LSPs are anchored at the GW node. In addition to the complexities related to integrating IP/MPLS functionality into a RAN BS, the GW in this proposal is a single point of failure and a potential source of congestion.

In [15] (MM-MPLS-MIPv6) an IPv6 specific solution is proposed that introduces MPLS into HMIPv6. The MPLS micro-mobile domain uses RSVP signaling to manage LSPs on a per-MN basis thus requiring all MPLS nodes to participate in mobility management. In the overlay model MPLS is used within the HMIPv6 environment purely for the traffic delivery purposes (instead of IPv6 GTP tunnels). The authors point out that the lack of integration between MPLS and Mobile IP results in protocol inefficiencies requiring removal and reinsertion of MPLS headers for every packet by the Mobile IP entities (MAP, HA). The integrated model avoids this by allowing a Mobile IP entity to directly access the MPLS forwarding base. This proposal however is essentially the same as HMIPv6 but with added MPLS capabilities.

### *E. Summary*

As can be seen from the description above there are numerous proposals for handling macro- and micro-mobility. However, a common requirement for all of them is the use of MIP. In addition, MIP HA or its derivatives (LERG, LEMA, GW, LMA) is a single entity that provides mobility support and therefore represents a central resource that is subject to survivability, capacity and scalability considerations. The MPLS-based proposals all concentrate on providing micro-mobility and act as extensions to Mobile IP. The MPLS micro-mobility schemes do not directly address optimal traffic delivery issues and in turn raise significant scalability concerns due to the need in most of them for flow-driven per-MN LSP management requiring every node in the MPLS domain to take part in the mobility management.

Throughout all of the considered related work none of the schemes provides a common control plane for scalable micro- and macro-mobility support for IPv4, IPv6, mobile hosts and mobile routers as well as the associated forwarding plane that is capable of optimal traffic delivery even for MN-to-MN communications. Table 1 provides a summary of the features of the presented mobility management architectures including H-MLBN.

TABLE I

Summary of existing architectures and comparison with H-MLBN

| Architecture | Mobility Type | Hierarchy Levels | Control Protocol | Forwarding Plane Protocol | IP Ver. | MN | Traffic Routing | MPLS Label Distribution | Mobile IP Required |
|---|---|---|---|---|---|---|---|---|---|
| MIPv4 | Macro | 2 | MIPv4 | IPv4,GRE, IP-IP | 4 | Host | Sub-optimal | N/A | Yes |
| ROMIPv4 | Macro | 2 | MIPv4 | IPv4,GRE, IP-IP | 4 | Host | Optimal[1] | N/A | Yes |
| NEMOv4 | Macro | 2 | MIPv4 | IPv4, GRE | 4 | Router | Sub-optimal | N/A | Yes |
| MIPv6 | Macro | 2 | MIPv6 | IPv6, GTP | 6 | Host | Sub-optimal, Optimal[1] | N/A | Yes |
| NEMOv6 | Macro | 2 | MIPv6 | IPv6, GTP | 6 | Router | Sub-optimal | N/A | Yes |
| PMIP | Macro[2] | 2 | MIPv6 | IPv6, GTP | 4, 6 | Host | Sub-optimal | N/A | Yes |
| RRMIPv4 | Micro | 3 | MIPv4 | IPv4,GRE, IP-IP | 4 | Host | Sub-optimal | N/A | Yes |
| HMIPv6 | Micro | 3 | MIPv6 | IPv6, GTP | 6 | Host | Sub-optimal, Optimal[1] | N/A | Yes |
| LMM-MPLS | Micro | 3 | MIPv4 RSVP | IPv4,GRE, IP-IP MPLS | 4 | Host | Sub-optimal | Flow | Yes |
| MM-MPLS | Micro | 3 | MIPv4 RSVP | IPv4,GRE, IP-IP MPLS | 4 | Host | Sub-optimal | Flow | Yes |
| MCM-MPLS | Micro | 3 | MIPv4 LDP | IPv4,GRE, IP-IP, MPLS | 4 | Host | Sub-optimal | Topology | Yes |
| RAN-MM-MPLS | Micro | 3 | MIPv4 RSVP | IPv4,GRE, IP-IP MPLS | 4 | Host | Sub-optimal | Flow | Yes |
| MM-MPLS-MIPv6 | Micro | 3 | MIPv6 RSVP | IPv4,GTP,MPLS | 6 | Host | Sub-optimal | Flow | Yes |
| H-MLBN | Both | 3 | MP-BGP | MPLS | 4, 6 | Both | Optimal | Topology | No |

[1] Requires support on CN, produces additional security requirements for CN, HA
[2] Supports limited form of micro-mobility

## III. Movement Model

### *A. Mobility Region Structure*

A Mobility Region (MR) consists of a number of RAN cells. Each cell is represented by a hexagon with radius $r$ (as shown in Fig. 2a. The height of a cell is $r_h = r\sqrt{3}/2$. The cells in the MR are arranged in rings as shown in Fig. 3. The size of the MR is determined by the number of rings $L$. Each cell in the MR is labeled as $S_i^l$, where $1 \le i \le N_l$ is the cell number in the ring and $1 \le l \le L$ is the ring number. The rings are counted from 0 with the cell $S_1^0$ in the center of the MR. The number of cells in a given ring can be expressed as $N_l = 6l,\ l = 1, 2, 3...L$. Thus, the total number of cells in the Mobility Region can be written as:

$$N = 1 + \sum_{l=1}^{L} 6l = 3L(L+1) + 1 \quad (1)$$

In the MR shown in Fig. 3 $L = 3$ and $N = 37$. The area of the MR can be found as follows:

$$A_r = \frac{3\sqrt{3}}{2} N r^2 \quad (2)$$

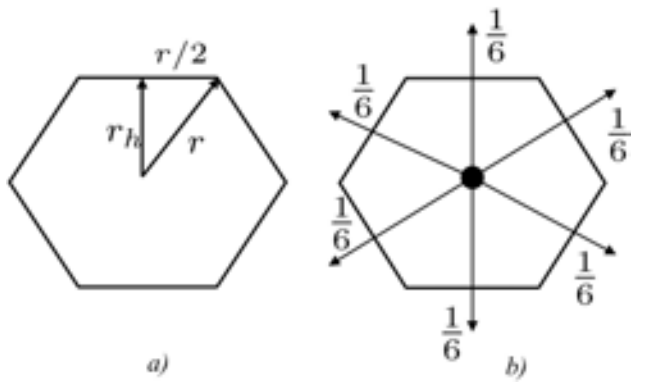

Figure 2. Cell radius (a), movement distribution (b).

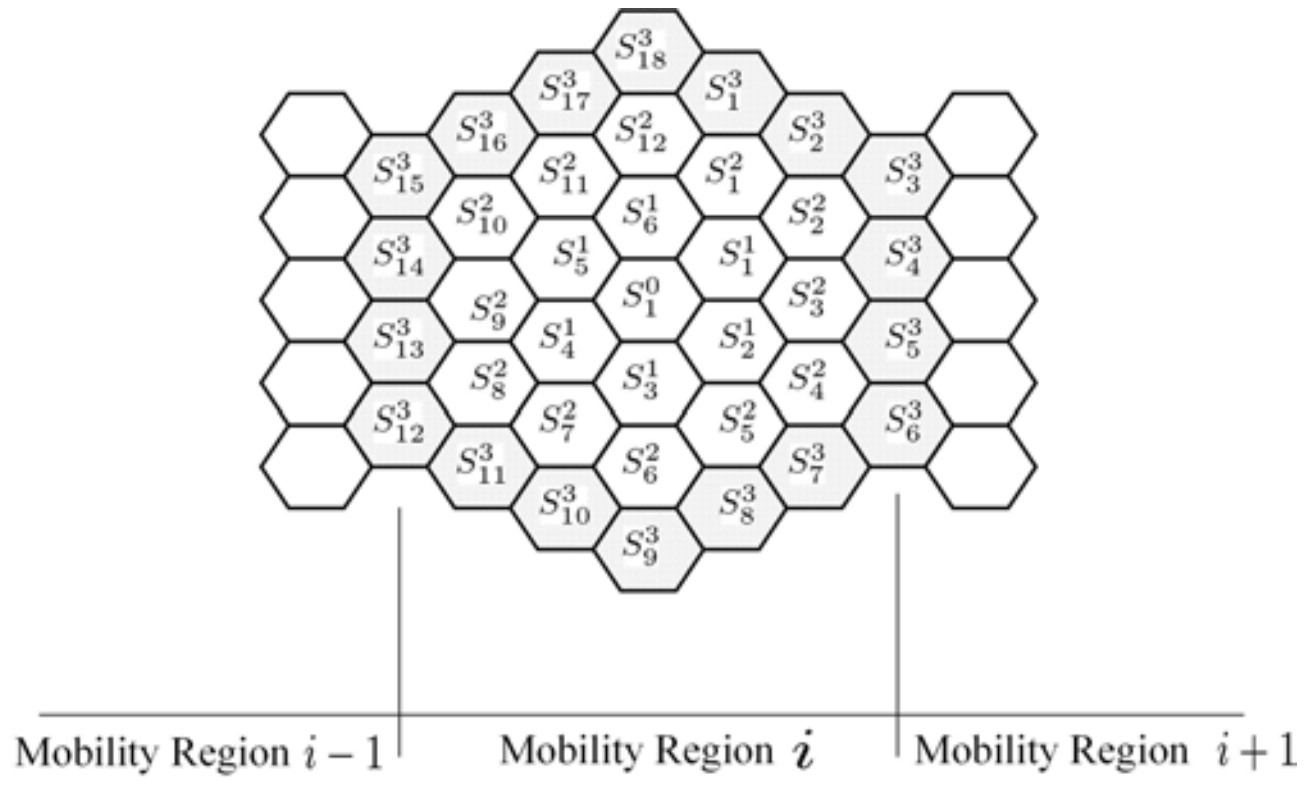

Figure 3. Mobility Region $L = 3,\ N = 37$.

### *B. Movement in the Mobility Region*

Consider a mobile node (MN) that is moving from cell to cell in the MR. As shown in Fig 2b once in a given cell the MN is allowed to move out of that cell in one of the six directions with equal probability. The dwelling time $T_c$ of the MN in a given cell is assumed to be exponentially distributed with parameter $\mu$ (similarly to [17]).

We define a moment in time when the MN transitions from a standby state to an active state as an "on-event" or visa versa as an "off-event". The active life time of the MN $T_l$ is defined as the time period between the on-event and the following off-event. The active life time may represent the duration of time that the MN is known to the network.

We assume that $T_l$ is exponentially distributed with parameter $\lambda$. Therefore the remaining active life time $T_l^r$ is also exponentially distributed with parameter $\lambda$.

Thus the probability $p_c$ that the MN will move through a given cell can be expressed as:

$$p_c = Pr(T_l^r > T_c) = \int_0^{\infty} [1 - F_{T_l^r}(t)] f_{T_c}(t) dt = \frac{\mu}{\mu + \lambda} \quad (3)$$

where $F_{T_i}(t)$ and $f_{T_i}(t)$ are a CDF and a PDF of the exponential distribution.

Consider a Life-to-Mobility Ratio (LMR) defined as:

$$\rho_c = \frac{E[T_c]}{E[T_l^r]} = \frac{\lambda}{\mu} \quad (4)$$

where $E[T_c]$ is the average cell dwelling time and $E[T_l^r]$ is the average remaining active life time. LMR reflects the average number of cell changes per an active life time. Then $p_c$ can be written as:

$$p_c = \frac{1}{1 + \rho_c} \quad (5)$$

Consider a connected graph $R(V,\ E)$ that is constructed by connecting the centers of the cells in the Mobility Region (Fig. 3) using the directions of possible movements of the MN from within a given cell (Fig. 2b). The cells of the ring $l = L$ (shown as shaded in Fig. 3) form the border cells of the MR. The graph $R(V,\ E)$ with each vertex labeled with its degree is shown in Fig 4.

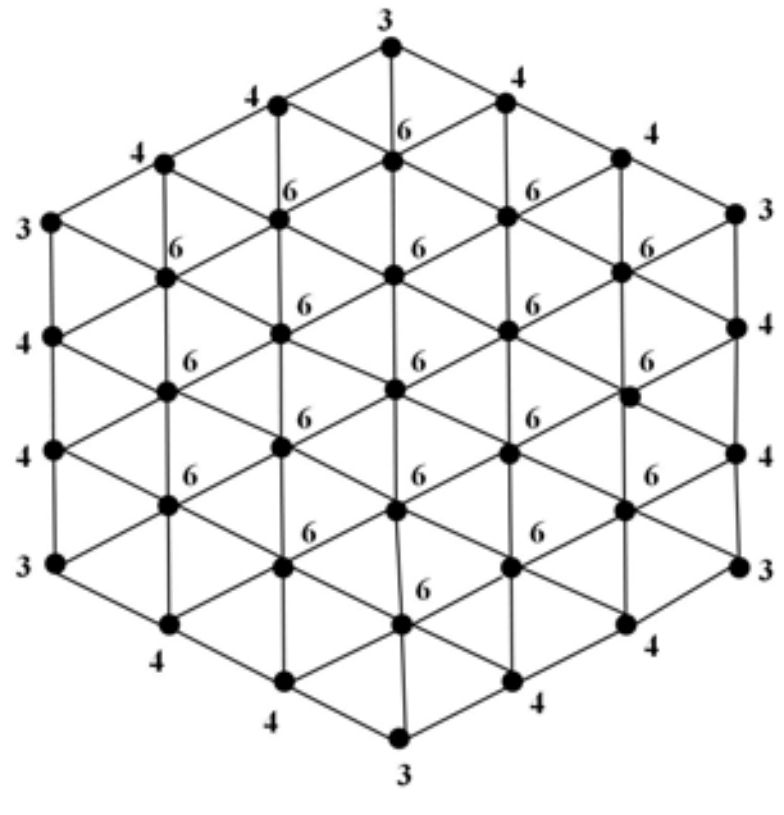

Figure 4. Connected graph representing the Mobility Region

The movements of the MN within the MR are modeled as a random walk on $R(V,\ E)$. This random walk can be represented by a Markov chain with the stationary distribution given by:

$$\pi_u = \frac{d_u}{2|E|}, \text{ where}$$

$d_u$ - is the degree of vertex $u$, and $|E|$ - is the total number of edges in $R(V,\ E)$. According to Fig. 4 the set of vertices of graph $R$ can be written as $V = V^I \bigcup V^E \bigcup V^C$, where $d_{u\in V^I} = 6$ (Internal vertices), $d_{u\in V^E} = 4$ (Edge vertices) and $d_{u\in V^C} = 3$ (Corner vertices). In addition:

$$|V^I| = 1 + \sum_{l=1}^{L-1} 6l = 3L(L-1) + 1$$
$$|V^E| = 6(L-1)$$
$$|V^C| = 6$$

The total number of edges in $R$ can be found using:

$$2|E| = |V^I|d_{u\in V^I} + |V^E|d_{u\in V^E} + |V^C|d_{u\in V^C} = 6L(3L+1)$$

Therefore the stationary probabilities are:

$$\pi^r_{u\in V^I} = \frac{1}{L(3L+1)}$$
$$\pi^r_{u\in V^E} = \frac{2}{3L(3L+1)}$$
$$\pi^r_{u\in V^C} = \frac{1}{2L(3L+1)}$$

We are interested in the probability of crossing of the boundary of a MR $p_{sr}$. This probability can be expressed as:

$$p_{sr} = \frac{1}{3}p_c\pi^r_{u\in V^E} + \frac{1}{2}p_c\pi^r_{u\in V^C} \approx \frac{1}{2L(3L+1)(1+\rho_c)} \quad (6)$$

**Proposition 1**. The number of RAN cell boundary crossings $M_{cb}$ of the MN that results in the MN exiting the MR (including the crossing of the MR boundary) is Geometrically distributed with parameter $p_{sr}$:

$$Pr(M_{cb} = k) = (1-p_{sr})^{k-1}p_{sr} \quad (7)$$

**Proof:** If we let the random walk on graph $R(V,E)$ representing the MR progress over time, then at any point in time chosen at random (a typical time) the probability that the MN could cross the boundary of $R(V,E)$ is $p_{sr}$. Consider a sequence of observation intervals (steps) starting at a typical point in time. There may be $(k-1)$ steps with probability $(1-p_{sr})$ each before the success on the $k_{th}$ step with probability $p_{sr}$ ■.

**Proposition 2.** The dwelling time (the time to exit) of the MN in the MR $T_r$ is distributed exponentially with parameter $\eta_r = \mu p_{sr}$.

**Proof:** The dwelling time in the MR can be expressed as:

$$T_r = \sum_{k=1}^{M_{cb}} T_{c_k}$$

which is a random sum of exponentially distributed i.i.d. random variables ($T_{c_k}$ - cell dwelling times) with the geometrically distributed number of terms $M_{cb}$.

Consider a random variable:

$$T_{r|M_{cb}=n} = \sum_{k=1}^{n} T_{c_k}$$

which is a deterministic sum of i.i.d. exponential random variables. This sum is Gamma-distributed with the conditional density function:

$$f_{T_r|M_{cb}=n}(t|n) = \frac{\mu^n}{(n-1)!}t^{n-1}e^{-\mu t}, \quad t \geq 0$$

From this the probability density function of $T_r$ is:

$$f_{T_r}(t) = \sum_{n=1}^{\infty} f_{T_r|M_{cb}=n}(t|n)f_{M_{cb}}(n)$$
$$= \sum_{n=1}^{\infty} \frac{\mu^n}{(n-1)!}t^{n-1}e^{-\mu t}(1-p_{sr})^{n-1}p_{sr}$$
$$= \mu p_{sr}e^{-\mu t}\sum_{n=1}^{\infty}\frac{(\mu(1-p_{sr})t)^{n-1}}{(n-1)!}$$
$$= \mu p_{sr}e^{-\mu t}e^{\mu(1-p_{sr})t} = \mu p_{sr}e^{-\mu p_{sr} t} \quad ■$$

The probability that the MN will move through a given MR during its remaining active life time is:

$$p_r = Pr(T^r_l > T_r) = \frac{\mu p_{sr}}{\mu p_{sr} + \lambda} = \frac{1}{1+\rho_c/p_{sr}} = \frac{1}{1+2\rho_c(1+\rho_c)L(3L+1)} \quad (8)$$

### C. *Mobility Area Structure*

MRs are grouped into Mobility Areas (MA). An MR with $L$ rings is approximated as a square of the same area as shown in Fig.5. The side of the approximating square is $a = \sqrt{\frac{3\sqrt{3}}{2}}R$, where $R = \frac{\sqrt{3}}{2}(2L+1)r$, and $r$ is the cell radius as shown in Fig. 2a

An MA is represented as a square with $M^2$ MRs, where $M > 1$. The structure of a MA is shown in Fig. 6a. Each MR internal to the MA has eight neighbors.

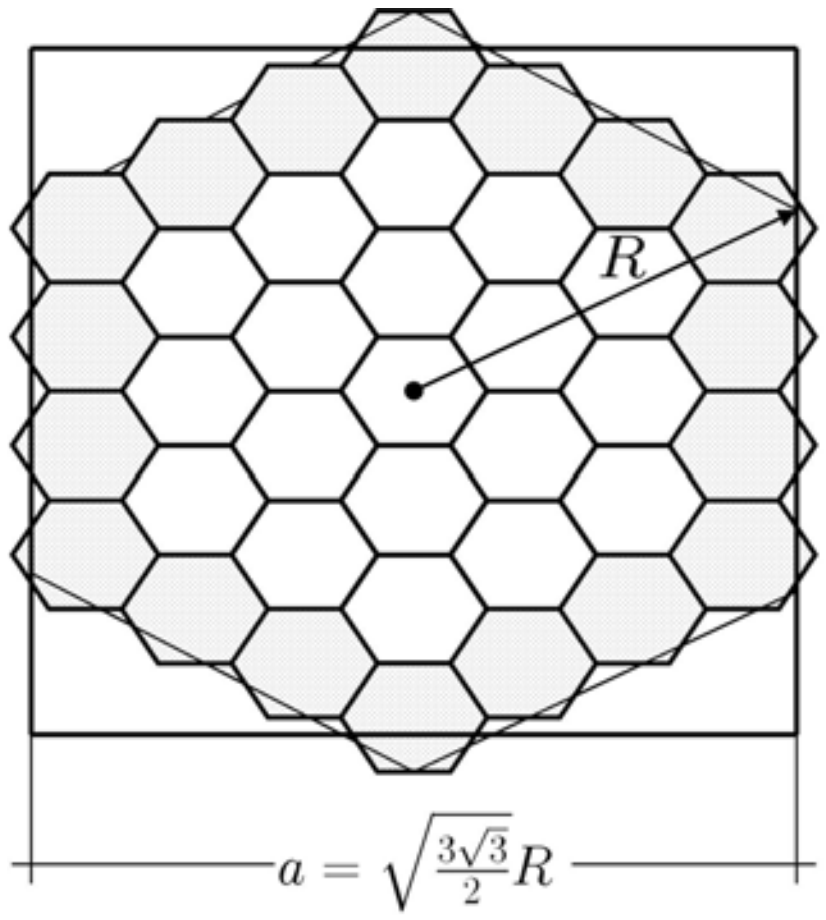

Figure 5. Approximating square for a Mobility Region

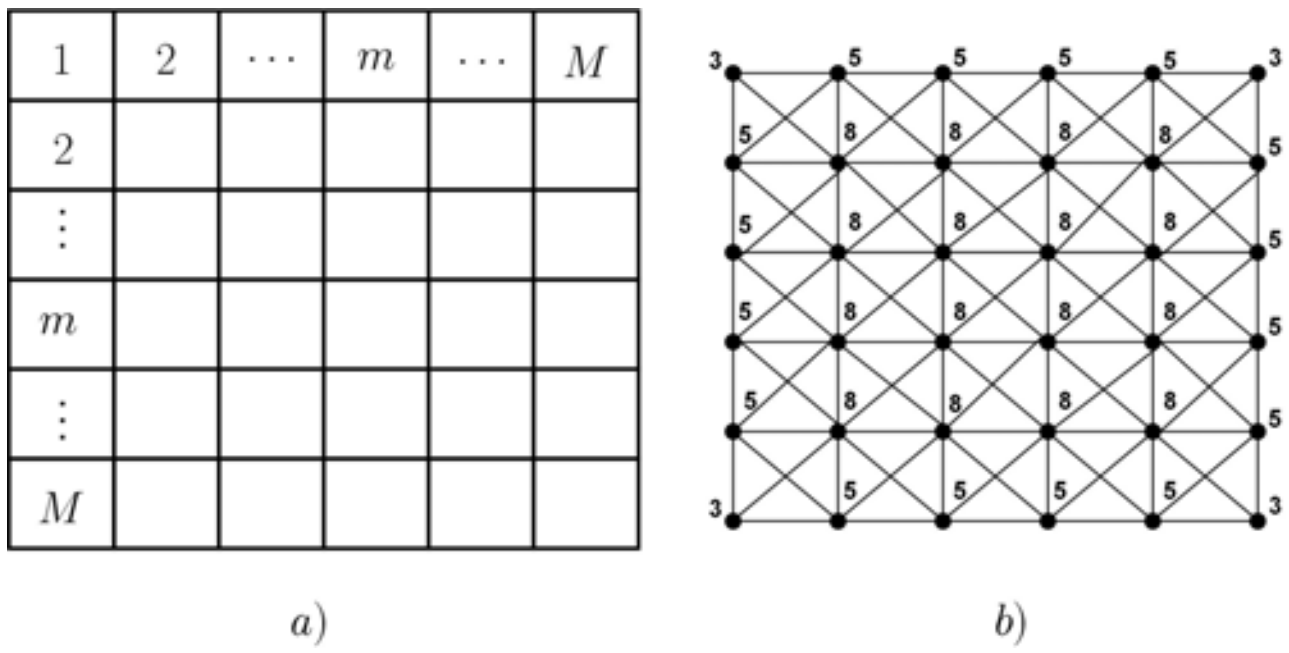

Figure 6. Mobility Area (a), Connected graph for Mobility Area (b)

The area of a MA $A_a$ can be expressed as follows:

$$A_a = \frac{9\sqrt{3}}{8}\left[Mr(2L+1)\right]^2 \approx 2\left[Mr(2L+1)\right]^2 \qquad (9)$$

*D. Movement in a Mobility Area*

Consider a connected graph $A(V, E)$ shown in Fig. 6b. Since, according to Proposition 2, the MN's dwelling time in a given MR is exponentially distributed, the random walk of the MN on $A(V, E)$ may be represented by a Markov chain. Proceeding as in the case of MR:

$$|V^I| = (M-2)^2$$
$$|V^E| = 4(M-2)$$
$$|V^C| = 4$$
$$2|E| = 4(2M-1)(M-1)$$

$$\pi^a_{u\in V^I} = \frac{2}{(2M-1)(M-1)}$$
$$\pi^a_{u\in V^E} = \frac{5}{4(2M-1)(M-1)}$$
$$\pi^a_{u\in V^C} = \frac{1}{(2M-1)(M-1)}$$

The probability of crossing of the boundary of a MA $p_{sa}$ is:

$$\begin{aligned} p_{sa} &= \frac{3}{8}p_r\pi^a_{u\in V^E} + \frac{5}{8}p_r\pi^a_{u\in V^C} \\ &= \frac{35p_r}{32(2M-1)(M-1)} \\ &\approx \frac{p_r}{(2M-1)(M-1)} \end{aligned} \qquad (10)$$

Using the same line of reasoning as in Propositions 1 and 2, the number of region transitions $M_{rb}$ of the MN that results in the MN exiting the MA is distributed as: $Pr(M_{rb} = r) = (1-p_{sa})^{r-1}p_{sa}$, and the dwelling time of the MN in the MA $T_a$ is distributed exponentially with parameter ($\eta_a = \mu p_{sr}p_{sa}$). The probability that the MN will move through a given MA during its remaining active life time can be expressed as:

$$p_a = Pr(T^r_l > T_a) = \frac{1}{1+\rho_c/(p_{sr}p_{sa})}$$
$$= \frac{1}{1+(2M-1)(M-1)\left\{\frac{E[T_r]}{E[T_c]} + E^2[T_c]Var[T_r]\right\}} \qquad (11)$$

*D. Movement During Active Life Time*

Consider the MN that is moving from one RAN cell to another, from one MR to another, and from one MA to another. We are interested in the distributions of the following random variables: the total number of RAN cell boundary crossings $C$ (cell, region or area), the number of region or area boundary crossings $R$, and the number of area crossings $A$ that the MN will experience during its active life time $T_l$.

**Proposition 3.** The probability $C(m)$ that the MN will experience exactly $m$ RAN cell crossings during its active life time is:

$$C(m) = Pr(C = m) = p_c^m(1-p_c),\ m = 0, 1, 2\ldots \qquad (12)$$

**Proof:** Similarly to [28], consider Fig. 7, where:
$T_l \propto exp(\lambda)$ - exponentially distributed active life time
$T^r_{l_i}$ - remaining active life time $i = 1, \ldots, m$
$T_{c_j} \propto exp(\mu)$ - cell dwelling time $j = 0, \ldots\ m$
$T^r_{c_0}$ - remaining dwelling time in cell 0

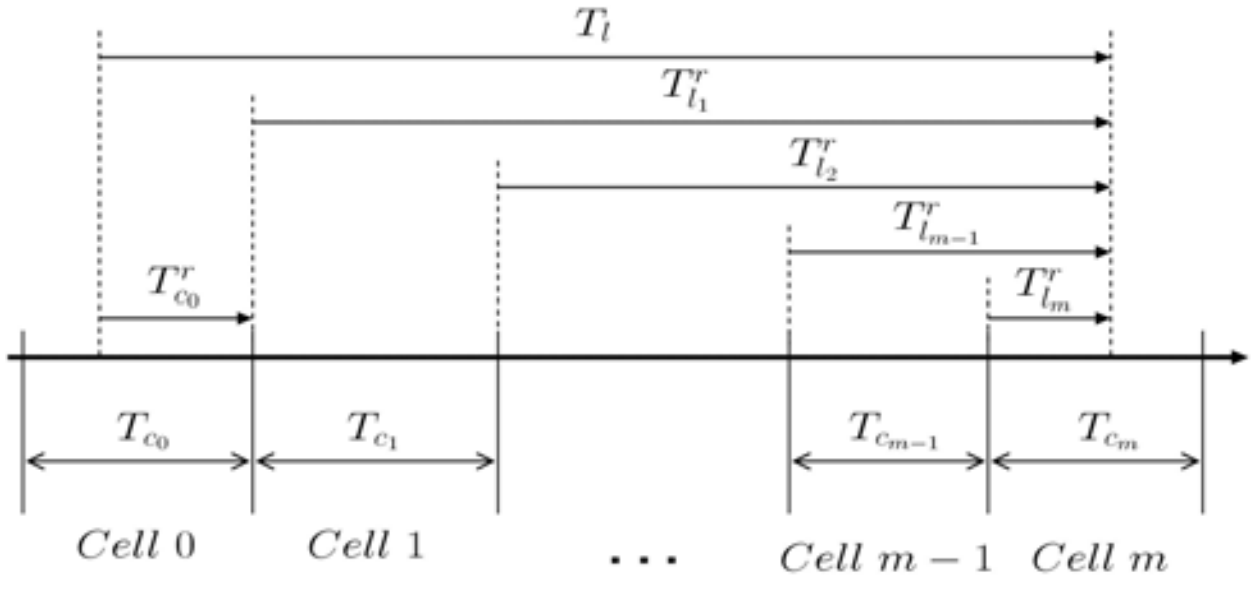

Figure 7. Mobility Area crossing diagram.

$C(m)$ may be expressed as:

$$Pr(T^r_{c_0}+T_{c_1}+\cdots+T_{c_{m-1}}<T_l\leq T^r_{c_0}+T_{c_1}+\cdots+T_{c_{m-1}}+T_{c_m})$$
$$=Pr(\Gamma_m<T_l\leq\Gamma_m+T_{c_m})$$
$$=Pr(T_l>\Gamma_m)[1-Pr(T_l>\Gamma_m+T_{c_m}|T_l>\Gamma_m)]$$
$$=Pr(T_l>\Gamma_m)[1-Pr(T_l>T_{c_m})]$$

Where:

$\Gamma_m = T^r_{c_0} + T_{c_1} + \cdots + T_{c_{m-1}}$ is a $Gamma(m,\mu)$-distributed random variable, and the second term in the product comes from the memoryless property of the exponential distribution.

$$Pr(T_l > \Gamma_m) = \int_0^\infty [1 - F_{T_l}(t)] f_{\Gamma_m}(t)dt$$
$$= \mu^m \int_0^\infty \frac{t^{m-1}}{(m-1)!} e^{-\lambda t} e^{-\mu t} dt = \mu^m F^*(\mu) \quad (13)$$
$$= \mu^m \left(\frac{1}{\mu+\lambda}\right)^m = p_c^m$$

where $F^*(\mu)$ is a Laplace transform of $f(t) = \frac{t^{m-1}}{(m-1)!}e^{-\lambda t}$ with $s = \mu$.

From (2) we also have that:

$$1 - Pr(T_l > T_{c_m}) = 1 - p_c \quad (14)$$

And using (13) and (14), (12) follows ■.

The average number and the variance of RAN cell crossings per MN's life time are:

$$E[C(m)] = \sum_{m=0}^{\infty} mC(m) = \frac{1}{\rho_c} \quad (15)$$

$$Var[C(m)] = \frac{1+\rho_c}{\rho_c^2} \quad (16)$$

Consider a conditional probability of the number of region or area crossings per an active life time given that the total number of cell crossings $C = m$:

$$R(r|C=m) = \binom{m}{r} p_r^r p_r^{m-r},\ m = r, r+1\ldots$$

The probability $R(r) = Pr(R = r)$ that the MN will experience $r$ mobility region or area crossings during its active life time is:

$$R(r) = \sum_{m=r}^{\infty} R(r|C=m)C(m)$$
$$= (1-p_c)\left[\frac{p_r}{1-p_r}\right]^r \sum_{m=r}^{\infty}\binom{m}{r}[p_c(1-p_r)]^m \quad (17)$$
$$= \frac{[1-p_c]}{[1-p_c(1-p_r)]}\left[\frac{p_c p_r}{1-p_c(1-p_r)}\right]^r$$

The average number and the variance of MR or MA crossings per MN's lifetime are computed as follows:

$$E[R(r)] = \sum_{r=1}^{\infty} rR(r) = \frac{p_r}{\rho_c} \quad (18)$$

$$Var[R(r)] = \left(1 + \frac{\rho_c}{p_r}\right)\frac{p_r^2}{\rho_c^2} \quad (19)$$

Further, consider a conditional probability of the number of MA crossings per an active life time given that $R = r$:

$$A(k|R=r) = \binom{r}{k} p_a^k p_a^{r-k},\ r = k, k+1\ldots$$

The probability $A(k) = Pr(A = k)$ that the MN will experience $k$ MA crossings during its active life time is:

$$A(k) = \sum_{r=k}^{\infty} A(k|R=r)R(r)$$
$$= \frac{[1-p_c]}{[1-p_c(1-p_r p_a)]}\left[\frac{p_c p_r p_a}{1-p_c(1-p_r p_a)}\right]^k \quad (20)$$

The average number and the variance of MA crossings per MN's lifetime are:

$$E[A(k)] = \sum_{k=1}^{\infty} kA(k) = \frac{p_r p_a}{\rho_c} \quad (21)$$

$$Var[A(k)] = \left(1 + \frac{\rho_c}{p_r p_a}\right)\frac{p_r^2 p_a^2}{\rho_c^2} \quad (22)$$

*E. Movement Between Mobility Areas*

Inter-Area movements of mobile devices are modeled as sequential transitions from one MA to another. MAs are structured as shown in Fig. 6 and represent large geographical network coverage areas. Clearly, due to the large coverage area of a MA (by construction), a given MN is expected to "survive" a finite number of $K$ area crossings before its active life expires. We choose $K$ such that for any small $\epsilon$:

$$Pr(T_l > KT_a) \leq \epsilon \implies K \geq \frac{\eta_a}{\lambda}\left(\frac{1}{\epsilon} - 1\right) \quad (23)$$

Consider a geography covered by $J$ MAs. The Inter-Area movements of MNs can be defined in the inter-area transition probability matrix $P_A$ given as follows:

$$P_{A_{ij}} = \begin{cases} A(|i-j|) & \text{for } |i-j| \le K \\ A(1) & \text{for } |i-j| = (J-1) \\ 0 & \text{otherwise} \end{cases} \tag{24}$$

Where, $i, j = 1 \dots J$, and $A(k)$ is defined in (20). In general the matrix $P_A$ may have the following structure:

$$P_A = \begin{bmatrix} 0 & A(1) & \dots & A(K) & 0 & \dots & 0 & A(1) \\ A(1) & 0 & A(1) & \dots & A(K) & 0 & \dots & 0 \\ \vdots & & & & & & & \vdots \\ A(K) & 0 & \dots & 0 & A(1) & 0 & A(1) & \dots \\ 0 & & & & & & & \\ \vdots & & & & & & & \vdots \\ 0 & & & & & & & \\ A(1) & 0 & \dots & 0 & A(K) & \dots & A(1) & 0 \end{bmatrix}_{J \times J}$$

The structure of $P_A$ reflects bi-directional circular Inter-MA movements of MN where the number of consecutive MA transitions is limited by $K$. The coverage area and the corresponding matrix $P_A$ for $J = 4$ and $K = 1$ are shown in Fig. 8.

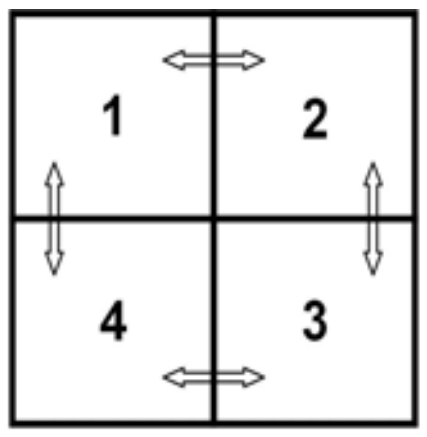

$a)$

$$P_A = \begin{bmatrix} 0 & A(1) & 0 & A(1) \\ A(1) & 0 & A(1) & 0 \\ 0 & A(1) & 0 & A(1) \\ A(1) & 0 & A(1) & 0 \end{bmatrix}_{4 \times 4}$$

$b)$

Figure 8. Sample coverage area with 4 Mobility Areas (a) and Inter-Area Transition Probability Matrix (b).

## IV. TRAFFIC MODEL

The traffic model presented in this section describes the forwarding plane traffic model and the control plane traffic model. The forwarding plane traffic model describes hop-by-hop forwarding behavior of the system for a communication path that is maintained between any two mobile nodes participating in a session. The control plane traffic model deals with the various types of network control events that result from the movements of mobile nodes.

### *A. Forwarding Plane Traffic Model*

Consider a pair of end nodes involved in a session. The two devices may be a mobile node and a fixed node or two mobile nodes. A packet sent from one MN to another will traverse a number of routers in the network. A hop is defined as a router hop that is counted when the packet crosses (or is switched by) a router (an H-MLBN node).

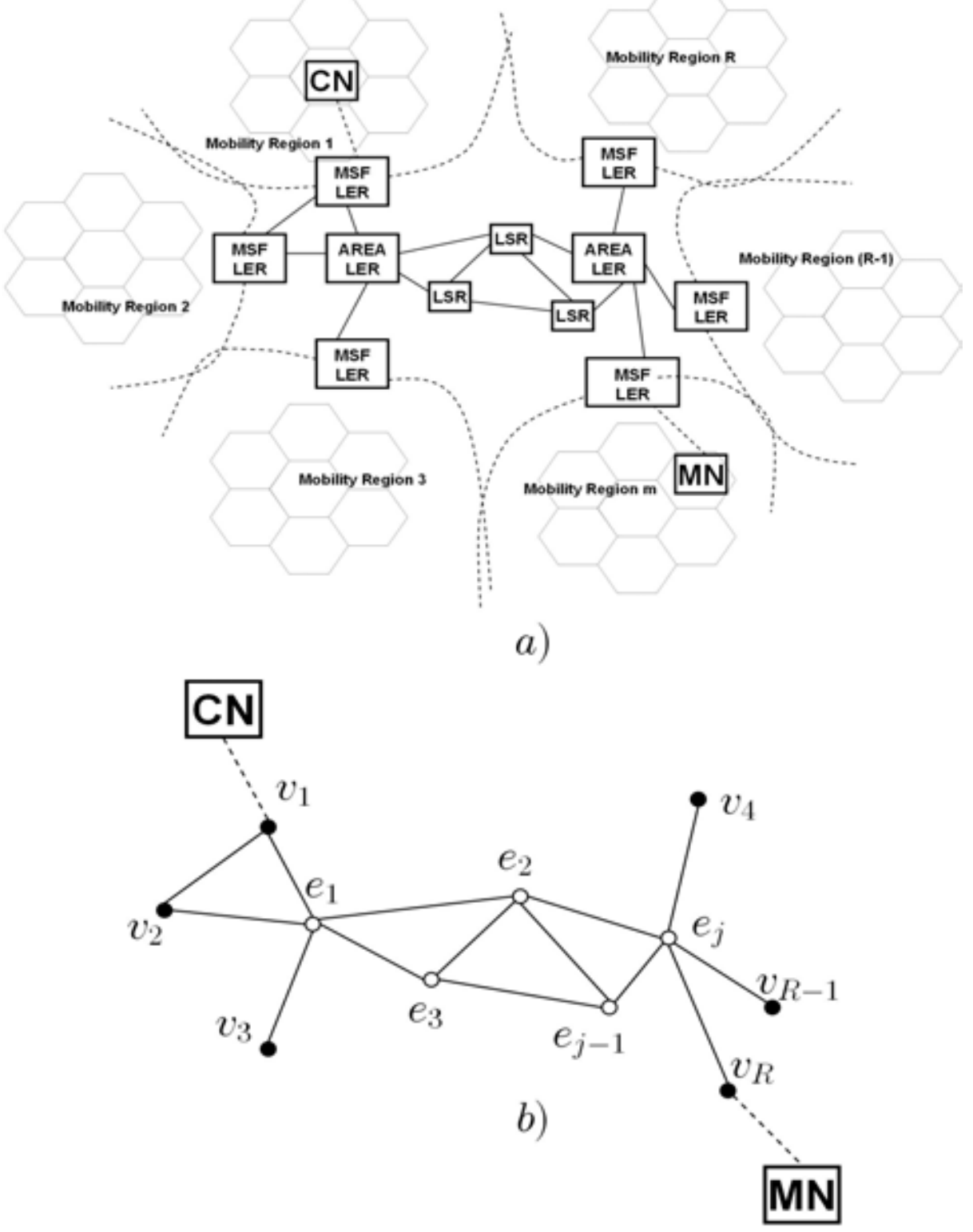

Figure 9. Sample H-MLBN with $D = 5$, a) network view, b) graph view.

Consider a network that is designed to allow a maximum network diameter of $D$ router hops between any two communicating nodes along an optimally routed path. Without loss of generality the optimally routed path is considered to be the shortest path in terms of the number of router hops. We assume a general network topology represented by a connected graph $G(V, E)$ where $V$ is a set of vertices and $E$ is a set of edges.

We associate a MR with each leaf vertex of $G$ and allow a mobile node to transition from one MR to another following a certain mobility pattern. Thus given a sub-set of leaf vertices $V^l \in V$ the total number of MRs can be defined as $R = |V^l|$. An example of this topology is shown in Fig. 9, where the nodes marked "LSR" and "AREA LER" correspond to internal vertices, and nodes marked "MSF LER" correspond to leaf vertices.

We then consider a mobile node that is moving between MRs and spending an average time of $1/\mu p_{sr}$ seconds in each region. As shown in Proposition 2 the dwelling time $T_r$ of a mobile node in a region is exponentially distributed with parameter $\eta_r = \mu p_{sr}$. We do not impose any specific restrictions on the pattern of movements of the mobile node, except to say that the probability of transitioning from one MR to another is $p_{sr}$ given in (6).

We first consider a moving MN communicating with a stationary CN. To represent relative locations of the communicating end nodes we select one leaf vertex $v_i \in V^l$of $G$ at random and designate it as the CN location and move the MN through all of the leaf vertices of $G$ (including $v_i$). At every move we calculate the value of $d_i$ which is the number of router hops between MN and CN along the optimally routed communications path. Clearly, $0 \le d_i \le D$. Given a general topology of graph $G$, at the completion of any regional move of the MN, $d_i$ may be represented by a uniformly distributed random variable: $d_i \backsim U[0 \quad D]$, where $q = Pr(d_i = j) = \frac{1}{D},\ 0 \le j \le D$. If this procedure is repeated for all possible locations for CN and MN, a process $\{X_s(t), t \ge 0\}$ could be formed with states $d_i, \quad i = 0, ..., D$ that represent the number of hops in the communication path between MN and CN. Note that the time that $X_s(t)$ spends in a given state is equal to $T_r$.

Thus $\{X_s(t), t \ge 0\}$ forms a Continuous Time Markov Chain (CTMC) with the state transition diagram shown in Fig. 10 below.

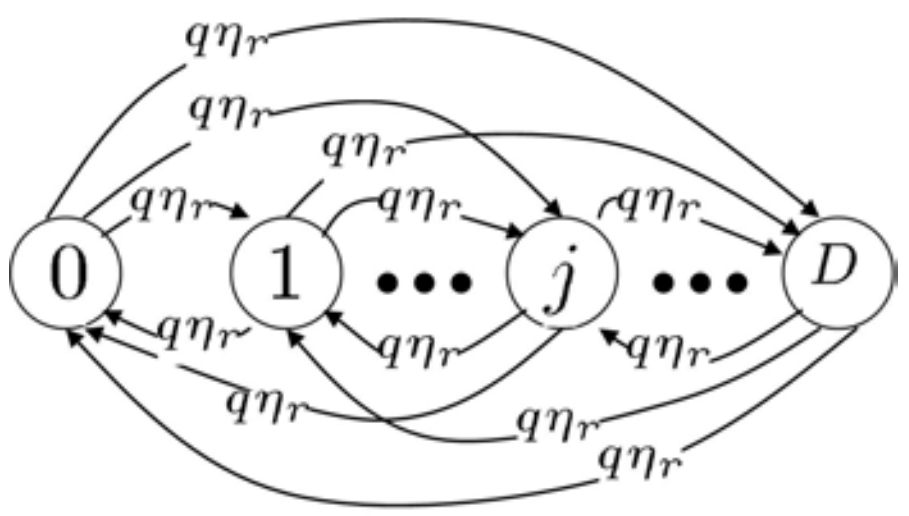

Figure 10. State transition diagram for the CTMC representing communications between MN and stationary CN where each state is the number of hops in the optimally routed communications paths.

Using global balance equations, the stationary distribution can be computed as follows:

$$\pi_0 D q\eta_r = \sum_{j=1}^{D} \pi_j q\eta_r = q\eta_r(1 - \pi_0)$$
$$\implies \pi_0 = \frac{1}{1+D} \tag{25}$$

and for $0 \le j \le D$:

$$\pi_j = \frac{1}{1+D} \tag{26}$$

Thus, we can find the expected value and variance of $X_s(t)$ as:

$$E[X_s(t)] = \sum_{j=0}^{D} j\pi_j = \frac{1}{(D+1)} \sum_{j=0}^{D} j = \frac{D}{2} \tag{27}$$

$$Var[X_s(t)] = \frac{D(D-2)}{12} \tag{28}$$

For the case of MN to MN communication, consider a process $\{X_m(t),\ t \ge 0\}$ with states $d_i, i = 0, \ldots, D$ as in the case of MN communicating to a fixed CN. Since each of the two communicating MNs may change its region, and the dwelling time of the MNs in a given region is exponential with parameter $\eta_r$, the time $X_m(t)$ spends in a given state is $min(T_{r_1},\ T_{r_2})$ , where $T_{r_1}$ and $T_{r_2}$ are the region dwelling times of the two communicating MNs. Therefore the time spent by$X_m(t)$ in a given state is also exponentially distributed but with parameter $2\eta_r$. The state transition diagram for $X_m(t)$ is the same as in Fig. 8 with $\eta_r$ replaced by $2\eta_r$. However, the stationary distribution, expected value and variance of $X_m(t)$ is exactly the same as for $X_s(t)$.

Expressions (27) and (28) give the average number and the variance of the number of hops in the optimally routed communication path between a moving MN and a fixed CN or between two moving MNs on a general network topology with a maximum diameter of $D$ router hops.

*B. Control Plane Traffic Model*

The movements of MNs in the coverage area result in various hand-off scenarios. In H-MLBN the following three hand-offs are defined: a) MSF-Local Hand-off – occurs every time MN crosses a RAN cell boundary within a given MR, b) Intra-Area Inter-MSF Hand-off – occurs every time MN crosses a MR boundary within a given MA, and c) Inter-Area Inter-MSF Hand-off – occurs when MN crosses a boundary of a MA. The control plane is responsible for performing network update procedures corresponding to each of the described hand-off types. For detailed information on the hand-off scenarios and the associated network updates please see [2].

Consider a MR $m$ with $N$ RAN cells and a coverage area $A_r$ as shown in (2). Let $\gamma_{r_m}$ be the rate of new active lives from region $m$ per unit time. We assume that new active lives are generated according to a Poisson process with rate $\gamma_{r_m} t$. Let $\gamma^I_{a_j}$ represent the rate of new active life origination internal to area $j$ per unit time. Assume that $\gamma^I_{a_j} t$ is the rate of the corresponding Poisson process. Clearly,

$$\gamma^I_{a_j} = \sum_{m=1}^{M^2} \gamma_{r_m} \tag{31}$$

Let $\gamma^T_{a_j}$ be the total transfer rate per unit time into area $j$ from all other areas. We assume that $\gamma^T_{a_j} t$ is the rate of a combined Poisson process. Given the Inter-Area transfer probability matrix $P_A$ (24) this rate can be expressed as:

$$\gamma_{a_j}^T = \sum_{i=1}^{J} \gamma_{a_i} P_{A_{ij}}, \quad j = 1 \dots J \qquad (32)$$

Where $\gamma_{a_j}$ is the total event rate (internal originations and transfers) incident on a MA. In addition, each of the active lives evolving in the network can be associated (marked) with a certain random number as follows:

$M_r$ - a number of MR boundary crossings.
$M_c$- a number of RAN cell boundary crossings.

The expectations of $M_r$ and $M_c$ may be computed as follows:

$$E[M_r] = E[R(r)] - E[A(k)] \qquad (33)$$

$$E[M_c] = E[C(m)] - E[R(r)] \qquad (34)$$

Note that $C(m) \geq R(r)$ represents the total number of boundary crossings (cell, region and area) during the active life of MN, $R(r) \geq A(k)$ represents the number of boundary crossings excluding the cell boundary crossings internal to a region or an area, and $A(k) \geq 0$ represents only the area boundary crossings. The relationship between the boundary crossing types is illustrated in Fig 11.

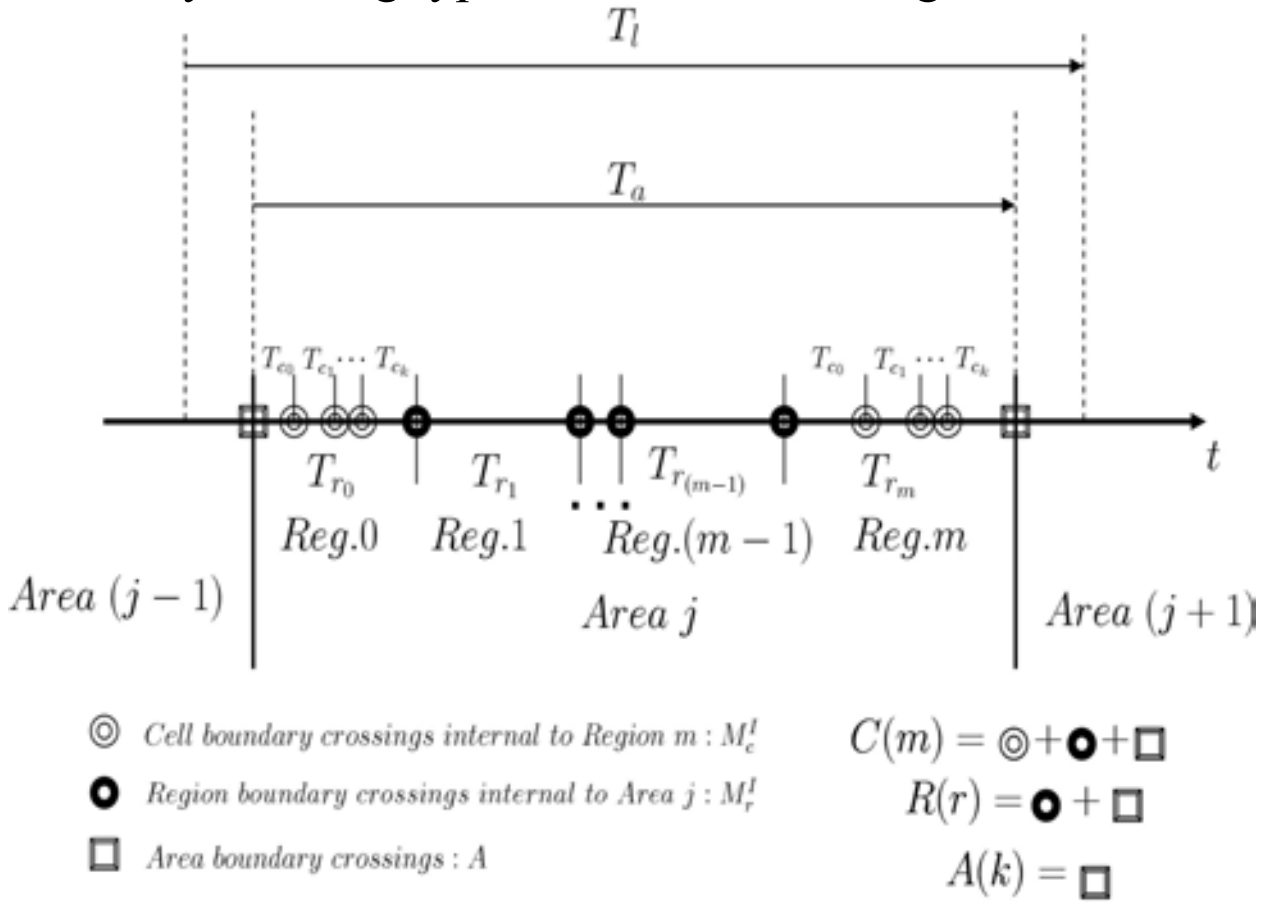

Figure 11. Relationship between boundary crossings: cell, region and area.

Note that each active life originated internally in an area or transferred into an area also produces a number of RAN cell and region boundary crossings internal to the area and represented by $M_c^I$ and $M_r^I$ respectively with expectations given by:

$$\begin{aligned} E[M_c^I] &= E[M_c]A(0) \\ E[M_r^I] &= E[M_r]A(0) \end{aligned} \qquad (35)$$

Where $A(0)$ is the probability that all boundary crossings are local to a MA, as shown in (20)

The total event rate incident on a MA $j$ may be expressed as follows:

$$\begin{cases} \gamma_{a_j} = \gamma_{a_j}^I + \sum_{i=1}^{J} \gamma_{a_i} P_{A_{ij}} \\ j = 1 \dots J \end{cases} \qquad (36)$$

Expression (36) gives a total number of events (origination and transfer) per unit time for each MA. Written in a matrix notation:

$$\Upsilon_a = \Upsilon_a^I + \mathbf{P}_A \Upsilon_a \qquad (37)$$

It has a unique solution:

$$\Upsilon_a = (\mathbf{I} - \mathbf{P}_A)^{-1} \Upsilon_a^I \qquad (38)$$

Where,
$\Upsilon_a^I = [\gamma_{a_1}^I, \gamma_{a_2}^I, \dots, \gamma_{a_J}^I]$ - a column vector of new active life origination rates per unit time for each Mobility Area.
$\Upsilon_a = [\gamma_{a_1}, \gamma_{a_2}, \dots, \gamma_{a_J}]$ - a column vector of total event rates per unit time in each Mobility Area.
$\mathbf{P}_A$ - a $J \times J$ inter-area transition probability matrix.
$\mathbf{I}$ - a $J \times J$ diagonal identity matrix.

Note that an event is either an origination of a new active life of a mobile node that occurs locally in a MA or a transfer of a mobile node from another MA during it's active life. We refer to these events as mobility events. Each such mobility event generates one or more network updates depending on how the original mobility event evolves during the active life of a mobile node. For example, a new active life originated in an area may produce multiple RAN cell and region boundary crossings internal to the area as well as the area boundary crossings. These "derivative" events are referred to as the network update events because they trigger appropriate network update procedures.

The total number of network update events that take place in a MA $j$ with a coverage area $A_a$ (9) per unit time may be expressed as follows:

$$\gamma_{a_j}^U = \gamma_{a_j}^{LM} + \gamma_{a_j}^{IM} + \gamma_{a_j}^{TL} + \gamma_{a_j}^{TI} + \gamma_{a_j}^{TA} \qquad (39)$$

Where,
$\gamma_{a_j}^{LM}$ - MSF-Local update rate due to new active lives originated in area $j$ per unit time.
$\gamma_{a_j}^{IM}$ - Intra-Area Inter-MSF update rate due to new active lives originated in area $j$ per unit time.
$\gamma_{a_j}^{TL}$ - MSF-Local update rate in area $j$ per unit time due to transfers of MNs during their active lives from other areas.
$\gamma_{a_j}^{TI}$ - Intra-Area Inter-MSF update rate per unit time in area $j$ due to transfers of MNs during their active lives from other areas.

$\gamma_{a_j}^{TA}$- Inter-Area Inter-MSF update rate per unit time in area $j$ due to transfers of MNs during their active lives from other areas.

The update event rate types are illustrated in Figure 12.

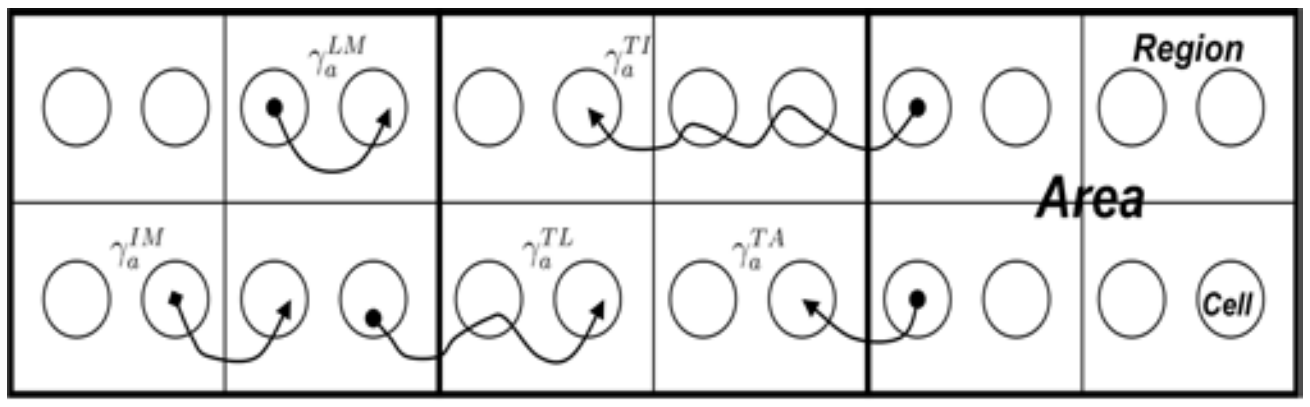

Figure 12. Network update event rate types.

Since each type of network update events involves different control plane procedures it is useful to consider them separately. Let,

$\gamma_{a_j}^{IA} = \gamma_{a_j}^{IM} + \gamma_{a_j}^{TI}$ - be a total rate of Intra-Area Inter-MSF update events per unit time in area $j$.

$\gamma_{a_j}^{LA} = \gamma_{a_j}^{LM} + \gamma_{a_j}^{TL}$ - be a total rate of MSF-Local update events per unit time in area $j$.

$\gamma_{a_j}^{TA}$ - be a total Inter-Area Inter-MSF event rate per unit time into area $j$.

Then:

$$\gamma_{a_j}^{IA} = \left[\gamma_{a_j}^{IM} + \gamma_{a_j}^{TI}\right] = \gamma_{a_j} E[M_r^I] \tag{40}$$

$$\gamma_{a_j}^{LA} = \left[\gamma_{a_j}^{LM} + \gamma_{a_j}^{TL}\right] = \gamma_{a_j} E[M_c^I] \tag{41}$$

$$\gamma_{a_j}^{TA} = \sum_{i=1}^{J} \gamma_{a_i} P_{A_{ij}} \tag{42}$$

Each of the expressions (40) – (42) can be written in the matrix notation:

$$\Upsilon_a^{IA} = \Upsilon_a E[M_r^I] \tag{43}$$

$$\Upsilon_a^{LA} = \Upsilon_a E[M_c^I] \tag{44}$$

$$\Upsilon_a^{TA} = \mathbf{P}_A \Upsilon_a \tag{45}$$

Where,

$\Upsilon_a^{IA} = [\gamma_{a_1}^{IA}, \gamma_{a_2}^{IA}, \ldots, \gamma_{a_J}^{IA}]$ - a column vector of Intra-Area Inter-MSF update rates per unit time in each MA.

$\Upsilon_a^{LA} = [\gamma_{a_1}^{LA}, \gamma_{a_2}^{LA}, \ldots, \gamma_{a_J}^{LA}]$ - a column vector of MSF-Local update rates per unit time in each MA.

$\Upsilon_a^{TA} = [\gamma_{a_1}^{TA}, \gamma_{a_2}^{TA}, \ldots, \gamma_{a_J}^{TA}]$ - a column vector of Inter-Area Inter-MSF update rates per unit time in each MA.

*C. Control Plane Processing Model*

The operation of the H-MLBN control plane is described in detail in [2]. The hand-off event rates derived above are the result of the movements of MNs in the network coverage area and are independent from the control plane processing. Each hand-off event generates a number of signaling messages that are processed by the appropriate network nodes. These nodes are:

LER – Label Edge Router. LER is responsible for performing MN registration, identifying the MN's movement type (from registration messaging) and executing a network update with the mobility binding information.

ALER – Area LER. ALER is a MPLS aggregation node serving a MA.

AMRR – Area Mobility Route Reflector is a control plane entity that is responsible for processing and distributing network updates.

The control plane processing model is used to derive the network update cost functions. The cost of updating the network consists of the signaling message delivery costs and processing costs. We use the following notation for the elements of the cost function:

$C_0$ - The cost of delivering the registration message between MN and LER.

$C_1$ - The cost of delivering a network update between LER and AMRR.

$C_2$ - The cost of delivering a network update between AMRR and ALER nodes.

$C_3$ - The cost of delivering a network update between two AMRR nodes.

$L_0$ - The processing cost of a local tracking (MSF-Local hand-off) operation at the LER node.

$L_1$ - Processing cost of creating a network update at LER (processing of MN registration, construction of a Mobility Binding message and management of Forwarding Information Base - FIB entries).

$L_2$ - Processing cost at AMRR (reflection of internal and external updates and Last Requestor List – LRL messages).

$L_3$ - Processing cost at ALER (processing and generation of internal and external updates and management of FIB entries).

$R_r$ - Serving rate of a radio link.

$R_w$ - Serving rate of a wireline link.

$d_r$ - Latency of a radio link.

$d_w$ - Latency of a wireline link.

$s_r$ - Average size of a registration message.

$s_u$ - Average size of a network update message.

$h_1$ - Average number of links between LER and AMRR within a given area.

$h_2$ - Average number of links between AMRR and ALER in a given area.

$h_3$ - Average number of links between two AMMRs in the network.

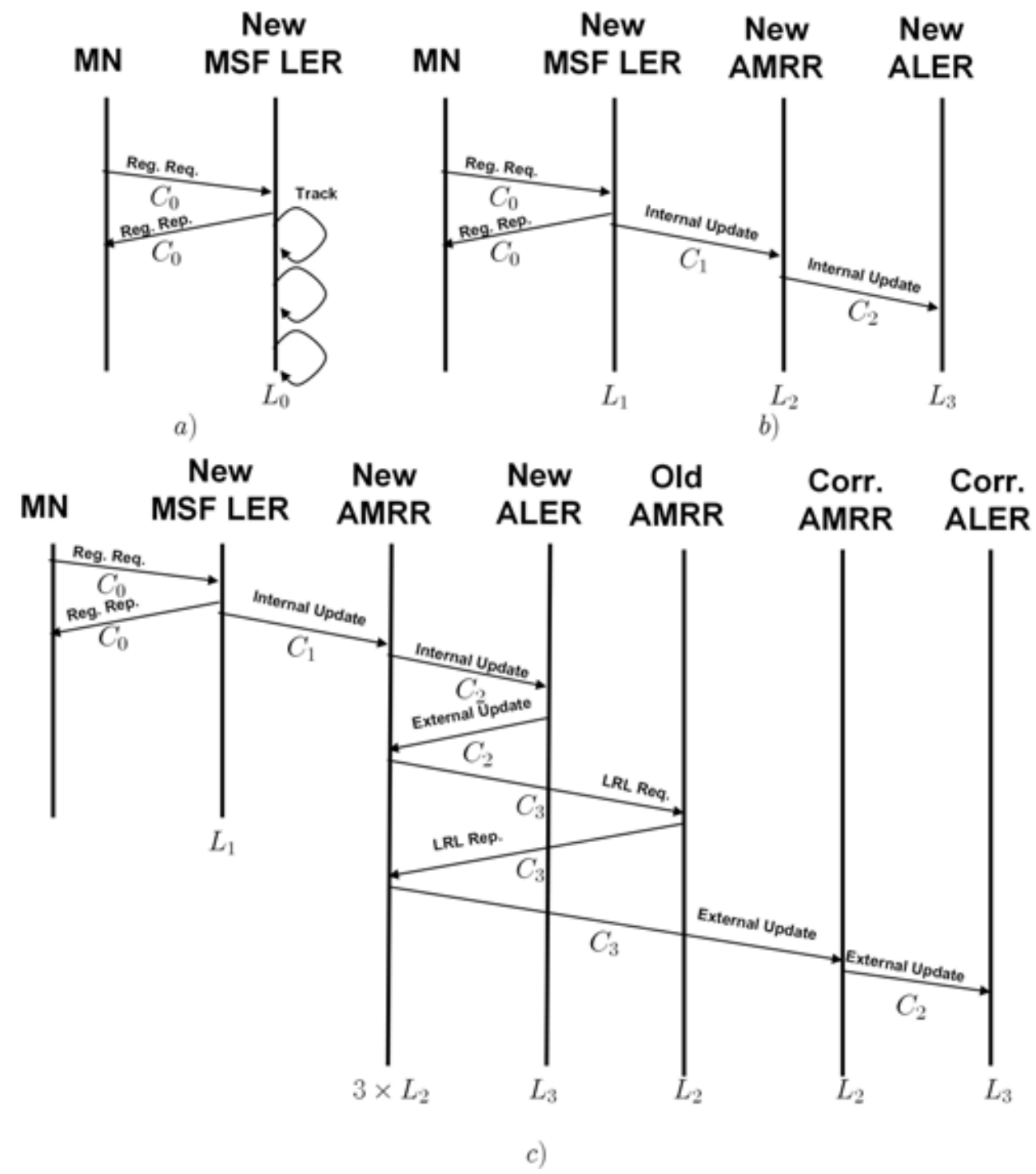

Figure 13. Control plane network update messaging: a) local tracking (MSF-Local hand-off), b) Intra-Area update (Intra-Area Inter-MSF hand-off), c) Inter-Area update (Inter-Area Inter-MSF hand-off).

Fig. 13 shows schematics of the network update procedures in H-MLBN. When MN's active life starts or when MN crosses the MA boundary or the MR boundary it performs a registration with a serving MSF LER and continues to transition between the RAN cells. While within a given MR each RAN cell boundary crossing by the MN results in the MSF-Local hand-off which is handled by the serving MSF LER by performing a local tracking operation similar to [9] (re-association of the MN's registration record with a new logical interface serving a given RAN cell). This is shown in Fig. 13a.

While within a given MA MN may transition between MRs. Each such transition (MR boundary crossing) results in an Intra-Area Inter-MSF hand-off. As shown in Fig. 13b this hand-off begins with the MN's registration with the new serving MSF LER, followed by the network update from the LER to the AMRR and reflection of that update to the ALER node. This sequence of messages results in ALER updating its Forwarding Information Base (FIB) record for the MN with a new current Mobility Label and the reconfiguration of the LSP to the MN's location via the new serving MSF LER.

When MN crosses the boundary of a MA an Inter-Area Inter-MSF hand-off is performed (Fig. 13c). This hand-off proceeds identically to the Intra-Area Inter-MSF hand-off up to and including the point in time when the new ALER node updates its FIB record for the MN. The new current Mobility Label assigned by the new ALER is communicated to the correspondent ALER nodes (nodes that serve sessions to the MN in question) via a sequence of messages between the new AMRR, old AMRR and the correspondent AMRR. The new AMRR identifies the old AMRR for the area from which the MN transitioned into the new area by exchanging Last Requestor List (LRL) messages with the old AMRR. The LRL returns the list of Area-IDs of the correspondent AMRRs. The new AMRR updates the correspondent AMRR with the new current Mobility Label for the MN, and the correspondent AMRR performs an update of the correspondent ALER within its area.

The signaling message delivery cost reflects the network link load induced by the signaling messages. It depends on the size of the message, the rate of messages, and the number of links (hops) the message needs to traverse. For a single signaling message:

$$\begin{cases} C_0 = s_r, & \text{if message} = \text{Registration} \\ C_i = h_i s_u, & \text{if message} = \text{Update}, \; i = 1,2,3 \end{cases} \quad (46)$$

Then, for a given MA, control plane message delivery costs for the three types of hand-offs are:

$$\begin{cases} C_{d_j}^{LA} = 2 s_r \gamma_{a_j}^{I} \\ C_{d_j}^{IA} = (2 s_r + s_u [h_1 + h_2]) \, \gamma_{a_j}^{IA} \\ C_{d_j}^{TA} = (2 s_r + s_u [h_1 + 3(h_2 + h_3)]) \, \gamma_{a_j}^{TA} \end{cases} \quad (47)$$

Where for $j = 1, \ldots, J$:

$C_{d_j}^{LA}$ - Control plane message delivery cost for MSF-Local hand-offs in area $j$ per unit time. Note that this cost is equal to the registration message delivery cost for the new MN active lives originated in the area. This is because after the initial registration the MSF-Local hand-offs do not require re-registrations.

$C_{d_j}^{IA}$ - Control plane message delivery cost for Intra-Area Inter-MSF hand-offs in area $j$ per unit time.

$C_{d_j}^{TA}$ - Control plane message delivery cost for Inter-Area Inter-MSF hand-offs in area $j$ per unit time.

The processing cost reflects the computational load induced by the signaling messages. Control plane messages are handled by network processors on the LER, ALER and AMRR nodes. The processing cost is proportional to the number of instructions required to process the information carried by the signaling messages and the rate of signaling messages. If $L_i$, $i = 0, \ldots, 3$ represent the number of computational instructions required to process respective hand-off event components on each of the corresponding network nodes, then the processing costs for the three hand-offs for a given mobility area expressed as a number of instructions per unit time are:

$$\begin{cases} C_{p_j}^{LA} = L_0 \gamma_{a_j}^{LA} \\ C_{p_j}^{IA} = (L_1 + L_2 + L_3) \, \gamma_{a_j}^{IA} \\ C_{p_j}^{TA} = (L_1 + 5L_2 + 2L_3) \, \gamma_{a_j}^{TA} \end{cases} \quad (48)$$

Note that it is expected that $L_1 \approx L_2 \approx L_3 \gg L_0$.

## V. Performance and Comparative Analysis

In this section key performance metrics for evaluating the operation of H-MLBN are developed. These metrics include the average number of network links in use (link count) between communicating mobile nodes, the average hand-off time for various hand-off scenarios and the network update costs. The same set of metrics is derived for Mobile IP based schemes for comparison.

### A. Link Count

The most interesting use case for developing the link count metric is the case of two moving mobile nodes communicating with each other during an active session. The link count is defined simply as a number of communication links utilized during a session between the two mobile nodes. More specifically, since each of the MNs must utilize the radio access and the associated wireline layer 2 grooming network links, these links are not included in the link count. Therefore the link count is the number of communication links between the mobile nodes excluding the access links on both sides (see Fig. 9b). Clearly, the link count is directly related to the router hop count (as defined in Section IV-A). Let $Z_{mlbn}$ denote the average link count in the H-MLBN environment and $D$ the maximum network diameter in terms of the router hops, then according to (27):

$$Z_{mlbn} = \left\lceil \frac{D}{2} - 1 \right\rceil, \qquad D > 1 \tag{49}$$

In general in the Mobile IP (MIP) based environments (including MIPv4 and MIPv6) a communications path between two MNs must involve at least one Home Agent (HA). In some cases, depending on the relative locations of MNs and the network coverage area two HAs must be involved. For practical reasons (such as packet filtering and other security policies) reverse tunneling is also employed where traffic to and from MN is always tunneled to the HA. It is important to note that HA nodes themselves are usually not part of the transport network but rather are attached to the transport nodes (routers) via separate links (home links). To illustrate this consider Fig. 14 below.

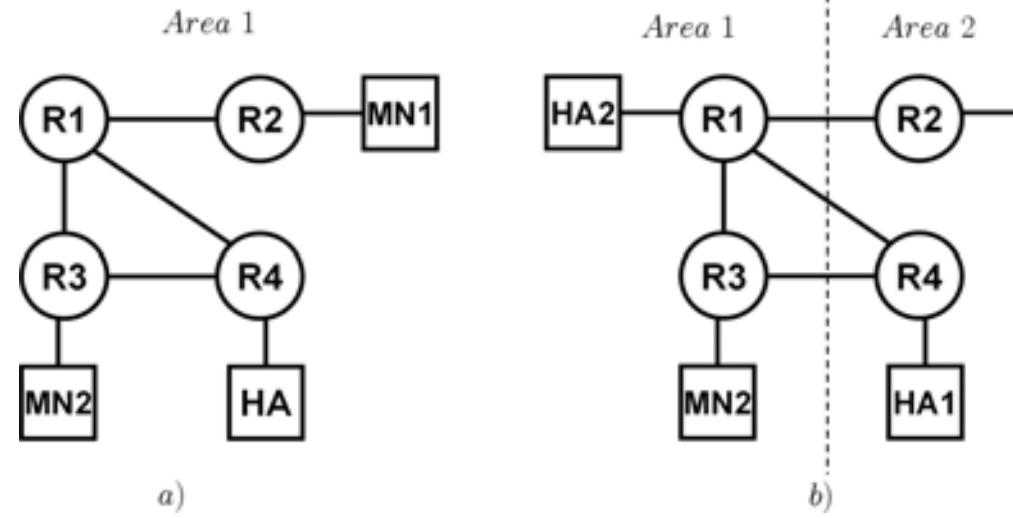

Figure 14. Communication path in a Mobile IP environment: a) single HA, b) two HAs.

The router network in Fig. 14 has a maximum diameter $D = 3$ router hops. In Fig. 14a both MNs are in the coverage area of the same HA and the communication paths are: $P_{12} = MN_1 \to MN_2 = \{R_2, R_1, R_4, HA, R_4, R_3\}$ and $P_{21} = MN_2 \to MN_1 = \{R_3, R_4, HA, R_4, R_1, R_2\}$. For both paths $P_{12}$ and $P_{21}$ the "triangular" hop count $D_t = 5$ (for comparison, the hop count for the optimal path $D_o = 3$).

In Fig. 14b $MN_1$ is served by $HA_1$ and $MN_2$ by $HA_2$. The communication paths between the MNs are:$P_{12} = \{R_2, R_1, R_4, HA_1, R_4, R_1, HA_2, R_1, R_3\}$ and $P_{21} = \{R_3, R_1, HA_2, R_1, R_4, HA_1, R_4, R_1, R_2\}$. For both paths $D_t = 7$ and $D_o = 3$.

The same traffic model as described in Section IV-A can be used for MIP-based scenarios. The state space of the CTMC (see Fig. 10) is then given as follows (assuming that the HA(s) is externally connected to the router network via a home link):

$$\begin{cases} \{0, 1, \ldots, 2D\} & \text{if single HA} \\ \{0, 1, \ldots, 3D\} & \text{if two HAs} \end{cases} \tag{50}$$

Therefore, according to (27) and including the home links:

$$Z_{mip} = \begin{cases} D & \text{if single HA} \\ \left\lceil \frac{1}{2}(3D + 1) \right\rceil & \text{if two HAs} \end{cases} \tag{51}$$

Note that the average link count $Z_{mip}$ is applicable to all MIP based schemes (including MPLS Micro-mobility) as traffic to MN is always routed via HA.

Link count may be used to express the network and user facing penalties caused by the triangular routing. Specifically, given the average traffic rate $R$ for a session between two MNs the excess network link utilization may be written as $U_l = (Z_{mip} - Z_{mlbn}) R$. This is a penalty that a network pays on a per-session basis for routing the session traffic using triangular path.

The user facing penalty for triangular routing may be defined as the additional delay and increased probability of packet loss due to the extra router hops that the session traffic needs to traverse on the sub-optimal path caused by triangular routing. Thus given the average packet delay $\delta$ and packet loss probability $p_l$ at every router hop, the additional delay $d$ and loss probability $l$ due to triangular routing may be expressed as $d = (Z_{mip} - Z_{mlbn}) \delta$ and $l = 1 - (1 - p_l)^{[Z_{mip} - Z_{mlbn}]}$.

### B. Hand-off Time

In H-MLBN, when a mobile device transitions between RAN cells and requires a hand-off it first needs to detect that the hand-off is required and then (if needed) perform a registration procedure with the new serving LER node, which in turn triggers the network update process. Thus the average hand-off time may be expressed as a sum of the following average time intervals:

$$T_{ho} = T_{hd} + T_{rr} + T_{nu} \tag{52}$$

Where:
$T_{ho}$ - Average hand-off time
$T_{hd}$ - Average hand-off detection time
$T_{rr}$ - Average re-registration time
$T_{nu}$ - Average network update time

The average hand-off detection time $T_{hd}$ is the time interval between the moment when MN last received data (a data packet or a heartbeat response packet) from the serving LER and the moment when MN determines that a re-registration procedure is required.

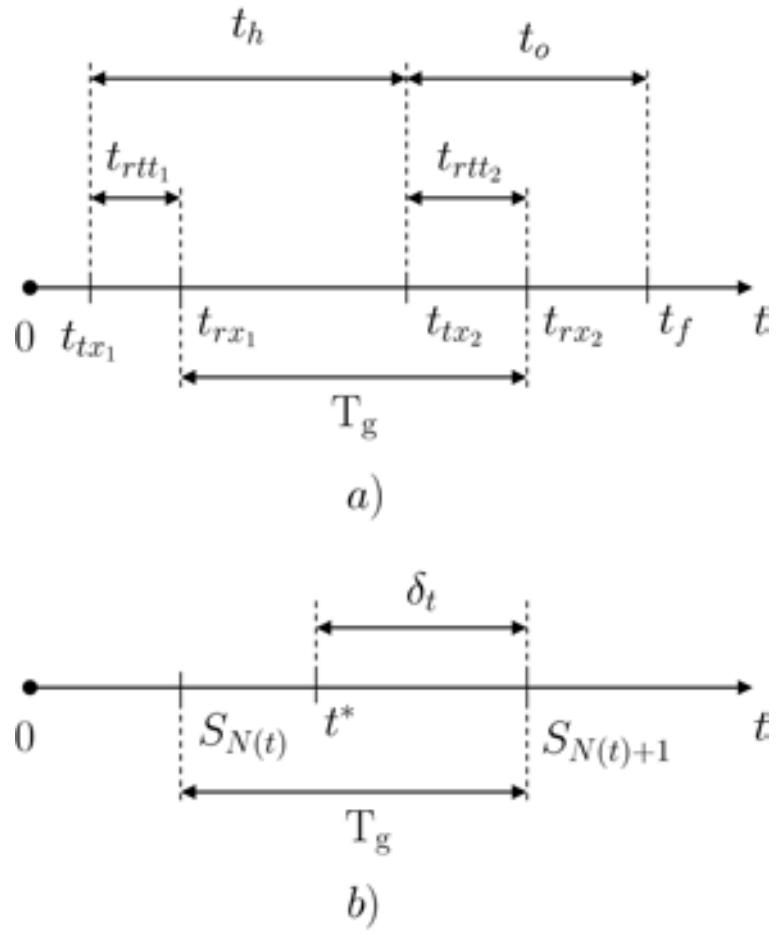

Figure 15. MN heartbeat and heartbeat timeout (a), hand-off detection (b).

Consider Fig. 15a. MN sends periodic heartbeat packets to the serving LER every $t_h$ deterministic time interval. If the first heartbeat packet is sent by MN at time $t_{tx_1}$ and the reply to it is received at time $t_{rx_1}$, and a second heartbeat packet is sent and the reply to it is received at times $t_{tx_2}$ and $t_{rx_2}$ respectively, then the time interval between the two successful heartbeat replies $\mathrm{T_g} = \mathrm{t_h} + (\mathrm{t_{rtt_2}} - \mathrm{t_{rtt_1}})$, where $t_{rtt}$ is the round trip time of the heartbeat request/reply transaction. If the heartbeat reply is not received by MN within the time interval $t_o$ after sending the request, MN detects loss of communication with the serving LER and initiates a re-registration procedure at time $t_f$. We assume that $\mathrm{T_g}$ is a uniformly distributed random variable: $\mathrm{T_g} \propto \mathrm{U}[\mathrm{t_h} - \mathrm{t_o}, \mathrm{t_h} + \mathrm{t_o}]$.

Consider a renewal process $\{N(t),\ t \geq 0\}$ with i.i.d. event inter-arrival times $\mathrm{T_g}$ and the times of the $n_{th}$ event $S_{N(t)=n} = \sum_{k=1}^{n} \mathrm{T_{g_k}}$. This process represents a sequence of successful heartbeat request/reply transactions between MN and the serving LER, where each event is the reception of a heartbeat reply in response to a transmitted heartbeat request as shown in Fig. 15b. Let $t^*$ represent a randomly chosen moment in time when MN movements result in a need for a hand-off. The hand-off detection time can be expressed as the time interval $\delta_t$ starting at $t^*$ and ending at the time when $\mathrm{T_g}$ expires. Therefore:

$$T_{hd} = E[\delta_t] = \frac{E[\mathrm{T_g^2}]}{2E[\mathrm{T_g}]} = \frac{t_h}{2} + \frac{t_o^2}{6t_h} \tag{53}$$

The average re-registration time $T_{rr}$ is the time that it takes the serving LER to process the registration request for MN. We assume that this time is approximately the same as the time that it takes the LER to execute a local tracking (MSF-Local hand-off) operation. Therefore (similar to [27]):

$$T_{rr} = \frac{L_0}{MIPS} \tag{54}$$

Where,
$L_0$ - Number of instruction executed by the Network Processor (NP) of the serving LER in order to process the registration request for MN.
$MIPS$ - Processing power (in instructions per second) available on (NP) to process the registration request.

The average network update time $T_{nu}$ depends on the type of hand-off that is being performed, and may be expressed as:

$$T_{nu} = \begin{cases} 0 & \text{MSF-Local} \\ \left(\frac{s_u}{R_w} + d_w\right)[h_1 + h_2] + \frac{L_1 + L_2 + L_3}{MIPS} & \text{Intra-Area} \\ \left(\frac{s_u}{R_w} + d_w\right)[h_1 + 3(h_2 + h_3)] + \frac{L_1 + 5L_2 + 2L_3}{MIPS} & \text{Inter-Area} \end{cases} \tag{55}$$

Another interesting performance parameter related to the average hand-off time is the average time spent by MN in hand-off processing during MN's active life. This metric may be expressed as follows:

$$T_{ho}^{l} = E[A(k)]T_{ho}^{Inter} + E[M_c^I]T_{ho}^{Local} + E[M_r^I]T_{ho}^{Intra} \tag{56}$$

Where,

$$T_{ho}^{Local} = T_{rr}$$
$$T_{ho}^{Intra} = T_{hd} + T_{rr} + T_{nu}^{Intra}$$
$$T_{ho}^{Inter} = T_{hd} + T_{rr} + T_{nu}^{Inter}$$

and $T_{nu}^{Intra}$ and $T_{nu}^{Inter}$ are given in (55).

In addition, we define hand-off intensity $\rho_h$ as a fraction of MN's active life that is spent during the hand-offs:

$$\rho_h = \frac{T^l_{ho}}{E[T_l]} = \lambda T^l_{ho} \quad (57)$$

In order to compare the hand-off performance of H-MLBN with that of the Mobile IP based schemes, the following assumptions are made:

- A RAN cell or a cell cluster is served by a MIP FA and corresponds to an IP sub-net in the foreign network. When MN transitions from one sub-net to another a re-registration with HA is required (in basic MIPv4/v6) or re-registration with RFA/MAP or their equivalents (in hierarchical schemes).
- H-MLBN Mobility Region corresponds to a regional coverage area by RFA or MAP in the hierarchical MIP.
- H-MLBN Mobility Area corresponds to a coverage area of a MIP HA.
- MN movement model is as described in Section III.
- Hand-off detection time is as described above.
- Processing cost of the MIP registration messages is approximately equal to $L_1$. Average registration message size is $s_r$, average binding update message size is $s_u$. Processing power of H-MLBN and MIP nodes is equal.
- Average number of network links between FA and HA is the same as $h_1$, between FA and RFA (or equivalent node) or between RFA and HA is $h_2$, and between any two HA's or any two RFA's is $h_3$. The wireline link serving rate is $R_w$.

All MIP based mobility management schemes may roughly be divided into basic and hierarchical schemes. Basic schemes include: MIPv4, MIPv6, PMIPv6 NEMOv4 and NEMOv6. We denote the collection of these protocols as B-MIP. Hierarchical schemes include: RRMIPv4 and HMIPv6 as well as the MPLS Micro-mobility schemes. Collectively, these protocols are denoted as H-MIP.

The inclusion of all MPLS-based schemes under the H-MIP category is based on a simple observation that all MPLS-based processing in support of mobility (LSP setup, teardown and redirection) is generally performed in addition to the MIP processing that is required to support the overall architecture of such solutions. Since these schemes act as extensions to hierarchical MIP the MIP processing represents the best case in terms of the overall system performance.

In addition, the hand-off type comparison between H-MLBN and B-MIP/H-MIP schemes along with the corresponding expectations is shown in Table II. Note that Inter-RFA hand-off is used to denote a hand-off within a coverage area of a RRMIPv4 RFA or HMIPv6 MAP. The Inter-HA hand-off applies to a situation when MN transitions to the new FA that is outside of the coverage area of MN's HA.

Although various proposals exist on how to possibly handle the inter-HA hand-offs, in general it is assumed that both B-MIP and H-MIP schemes require MN to establish a new registration with a new HA. Note that this registration is possible only when MN's Home Address is assigned dynamically by the HAs and that the active session between MN and CN cannot be maintained (as the Home Address of MN must change).

TABLE II
Hand-off Type Mapping

| H-MLBN | | H-MIP | | B-MIP | |
|---|---|---|---|---|---|
| **Type** | **Exp.** | **Type** | **Exp.** | **Type** | **Exp.** |
| **MSF-Local** | $E[M^I_c]$ | **Intra-RFA** | $E[M^I_c]$ | **Intra-HA** | $E[M^I_c] + E[M^I_r]$ |
| **Inter-MSF Intra-Area** | $E[M^I_r]$ | **Inter-RFA** | $E[M^I_r]$ | **Intra-HA** | $E[M^I_c] + E[M^I_r]$ |
| **Inter-MSF Inter-Area** | $E[A(k)]$ | **Inter-HA** | $E[A(k)]$ | **Inter-HA** | $E[A(k)]$ |

For the MIP based schemes the average hand-off time may be expressed as follows:

$$T_{ho_{(h-mip)}} = \begin{cases} T_{hd} + 2\left(\frac{S_u}{R_w} + d_w\right) h_2 + \frac{4L_1}{MIPS} & \text{Intra-RFA} \\ T_{hd} + 4\left(\frac{s_u}{R_w} + d_w\right) h_2 + \frac{6L_1}{MIPS} & \text{Inter-RFA} \\ T_{hd} + 2T_{st} + 4\left(\frac{S_u}{R_w} + d_w\right) h_2 + \frac{6L_1}{MIPS} & \text{Inter-HA} \end{cases} \quad (58)$$

$$T_{ho_{(b-mip)}} = \begin{cases} T_{hd} + 2\left(\frac{S_u}{R_w} + d_w\right) h_1 + \frac{4L_1}{MIPS} & \text{Intra-HA} \\ T_{hd} + 2T_{st} + 2\left(\frac{S_u}{R_w} + d_w\right) h_1 + \frac{4L_1}{MIPS} & \text{Inter-HA} \end{cases} \quad (59)$$

In addition:

$$T^l_{ho_{(h-mip)}} = E[A(k)] T^{Inter-HA}_{ho_{(h-mip)}} + E[M^I_c] T^{Intra-RFA}_{ho_{(h-mip)}} + E[M^I_r] T^{Inter-RFA}_{ho_{(h-mip)}} \quad (60)$$

$$T^l_{ho_{(b-mip)}} = E[A(k)] T^{Inter-HA}_{ho_{(b-mip)}} + \left(E[M^I_r] + E[M^I_c]\right) T^{Intra-HA}_{ho_{(b-mip)}} \quad (61)$$

Where,

$T_{st}$ - A session tear-down interval (an average time that it takes to terminate or re-establish the application session).

The hand-off intensity for the MIP based schemes is:

$$\rho_{h_{(h-mip)}} = \frac{T^l_{ho_{(h-mip)}}}{E[T_l]} = \lambda T^l_{ho_{(h-mip)}} \quad (62)$$

$$\rho_{h_{(b-mip)}} = \frac{T^l_{ho_{(b-mip)}}}{E[T_l]} = \lambda T^l_{ho_{(b-mip)}} \quad (63)$$

Note that the hand-off time analysis does not take the "fast versions" of MIPv4 [24] and MIPv6 [25] into consideration and only compares the performance of the layer 3 based hand-offs.

### *C. Registration and Network Update Cost*

For H-MLBN the registration and network update cost functions related to the three hand-off types are given in (47) and (48).

For MIP we again consider the B-MIP and H-MIP schemes as described above as well as follow the same assumptions. Specifically, based on the movement and traffic models, the derived hand-off event rates are re-used. Table III provides a mapping between the H-MLBN hand-off event rates and their interpretation for the MIP based environments.

TABLE III
Event Rate Mapping

| H-MLBN | | H-MIP | | B-MIP | |
|---|---|---|---|---|---|
| **Type** | **Rate** | **Type** | **Rate** | **Type** | **Rate** |
| **MSF-Local** | $\Upsilon_a^{LA}$, $\Upsilon_a^{I}$ | **Intra-RFA** | $\Upsilon_a^{LA}$, $\Upsilon_a^{I}$ | **Intra-HA** | $\Upsilon_a^{LA}$, $\Upsilon_a^{I}$ |
| **Inter-MSF Intra-Area** | $\Upsilon_a^{IA}$ | **Inter-RFA** | $\Upsilon_a^{IA}$ | **Intra-HA** | $\Upsilon_a^{LA}$, $\Upsilon_a^{I}$ |
| **Inter-MSF Inter-Area** | $\Upsilon_a^{TA}$ | **Inter-HA** | $\Upsilon_a^{TA}$ | **Inter-HA** | $\Upsilon_a^{TA}$ |

The message delivery costs for a given HA coverage area $j$ per unit area per unit time for H-MIP and B-MIP schemes may be expressed as follows:

$$\begin{cases} C_{d_j}^{LA(h-mip)} = 2\left(s_r + 2s_u h_2\right)\gamma_{a_j}^{I} + 2\left(s_r + s_u h_2\right)\gamma_{a_j}^{LA} \\ C_{d_j}^{IA(h-mip)} = 2\left(s_r + s_u h_2\right)\gamma_{a_j}^{IA} \\ C_{d_j}^{TA(h-mip)} = 2\left(s_r + 2s_u h_2\right)\gamma_{a_j}^{TA} \end{cases} \quad (64)$$

$$\begin{cases} C_{d_j}^{LA(b-mip)} = 2\left(s_r + s_u h_1\right)\left(\gamma_{a_j}^{I} + \gamma_{a_j}^{LA}\right) \\ C_{d_j}^{TA(b-mip)} = 2\left(s_r + s_u h_1\right)\gamma_{a_j}^{TA} \end{cases} \quad (65)$$

The associated message processing costs are:

$$\begin{cases} C_{p_j}^{LA(h-mip)} = 2L_1\left(3\gamma_{a_j}^{I} + 2\gamma_{a_j}^{LA}\right) \\ C_{p_j}^{IA(h-mip)} = 6L_1\gamma_{a_j}^{IA} \\ C_{p_j}^{TA(h-mip)} = 6L_1\gamma_{a_j}^{TA} \end{cases} \quad (66)$$

$$\begin{cases} C_{p_j}^{LA(b-mip)} = 4L_1\left(\gamma_{a_j}^{I} + \gamma_{a_j}^{LA}\right) \\ C_{p_j}^{TA(b-mip)} = 4L_1\gamma_{a_j}^{TA} \end{cases} \quad (67)$$

Clearly control plane costs depend on the mobility level, and as the mobility level increases the costs are expected to increase. It is also expected that differences in control plane costs of various architectures will increase as the mobility level increases. On the other hand there exists a low enough mobility level beyond which the differences in control plane costs of various architectures become insignificant.

We define a Minimum Event Rate of Significance $\gamma_{a_j}^{MERS}$ as the event rate corresponding to $\rho_c = 1$ or equivalently $p_c = 1/2$. This event rate and the corresponding network update rates may be considered as baseline rates relative to which the control plane costs associated with higher mobility levels are computed and compared.

## VI. PERFORMANCE AND NUMERICAL RESULTS

To illustrate the operation of H-MLBN the following system parameters are used:

TABLE IV
Numerical Parameters

| Parameter | Value | Parameter | Value |
|---|---|---|---|
| $L$ | 4 | $s_u$ | 512 bytes |
| $N$ | 61 | $s_r$ | 256 bytes |
| $M$ | 5 | $d_w$ | 2 msec |
| $r$ | 5 km | $h_1$ | 4 |
| $T_l$ | 3600 sec | $h_2$ | 2 |
| $J$ | 10 | $h_3$ | 6 |
| $\rho_c$ | 0.01 - 10 | $R_w$ | 100 Mbps |
| $t_h$ | 10 sec | $T_{st}$ | 1000 msec |
| $t_o$ | 3 sec | $R$ | 64 Kbps |
| $MIPS$ | $10^6$ | $\delta$ | 5 msec |
| $L_0$ | 10 | $p_l$ | 0.005 |
| $L_1 = L_2 = L_3$ | 100 | $\gamma_{a_j}^{I}$ | $10^2$ |
| $\epsilon$ | $10^{-3}$ | $D$ | 20 |

With the above parameters, the coverage areas for a cell, a mobility region and a mobility area are 65, 3,962 and 99,052 km$^2$ respectively, with 61 cells per region and 25 regions per area. A network of 10 mobility areas is considered. The analytical results developed in Section V are verified by a discrete event simulation tracking the movements of a mobile device on a predefined geographical coverage grid and recording the relative frequencies of the boundary crossing events and the corresponding network hand-off events. A similar simulation is performed for the forwarding plane where the number of hops between the moving mobile devices is tracked on the same coverage grid with varying maximum network diameters.

Fig. 16 shows the relationship between the LMR $\rho_c$ (4) and the estimated speed of mobile nodes given a typical RAN cell radius $r$ of 5 km and the average active life time $T_l$ of 3,600 seconds. The four chosen representative $\rho_c$ values of 0.01, 0.1, 1 and 10 correspond to the following mobility levels respectively: very high, high, low, very low (e.g. plain, train or car, pedestrian, near stationary).

Fig. 17 depicts the average number of cell, region and area boundary crossings per active life time of a mobile node for different values of $\rho_c$. As can be seen, for high, low and very low mobility levels the number of mobility area boundary crossings is expected to be near zero, the number of region boundary crossings is also expected to be near zero for low and very low mobility levels. For the very high mobility level a mobile node during its active life time is expected to cross approximately 100 RAN cell boundaries (including the region and area boundaries). For the high mobility level the corresponding expectation is approximately 15. Figs. 18 a, b and c show the distributions of $C(m)$, $R(r)$ and $A(k)$ given in (12), (17) and (20) respectively and their realizations obtained via the simulation for different values of $\rho_c$.

Fig. 19a shows the link count (the average number of network links) between two communicating moving mobile nodes for the cases of using H-MLBN and Mobile IP with one and two Home Agents serving the mobile nodes. As can be seen the use of H-MLBN provides the lowest link count that corresponds to the optimal path (in terms of the router hops) between the MNs. The link count resulting from the use of Mobile IP with a single HA serving the two MNs is on average twice as greater than that of H-MLBN and the link count in the case of Mobile IP with two HAs serving the communicating MNs is four times greater than that of H-MLBN.

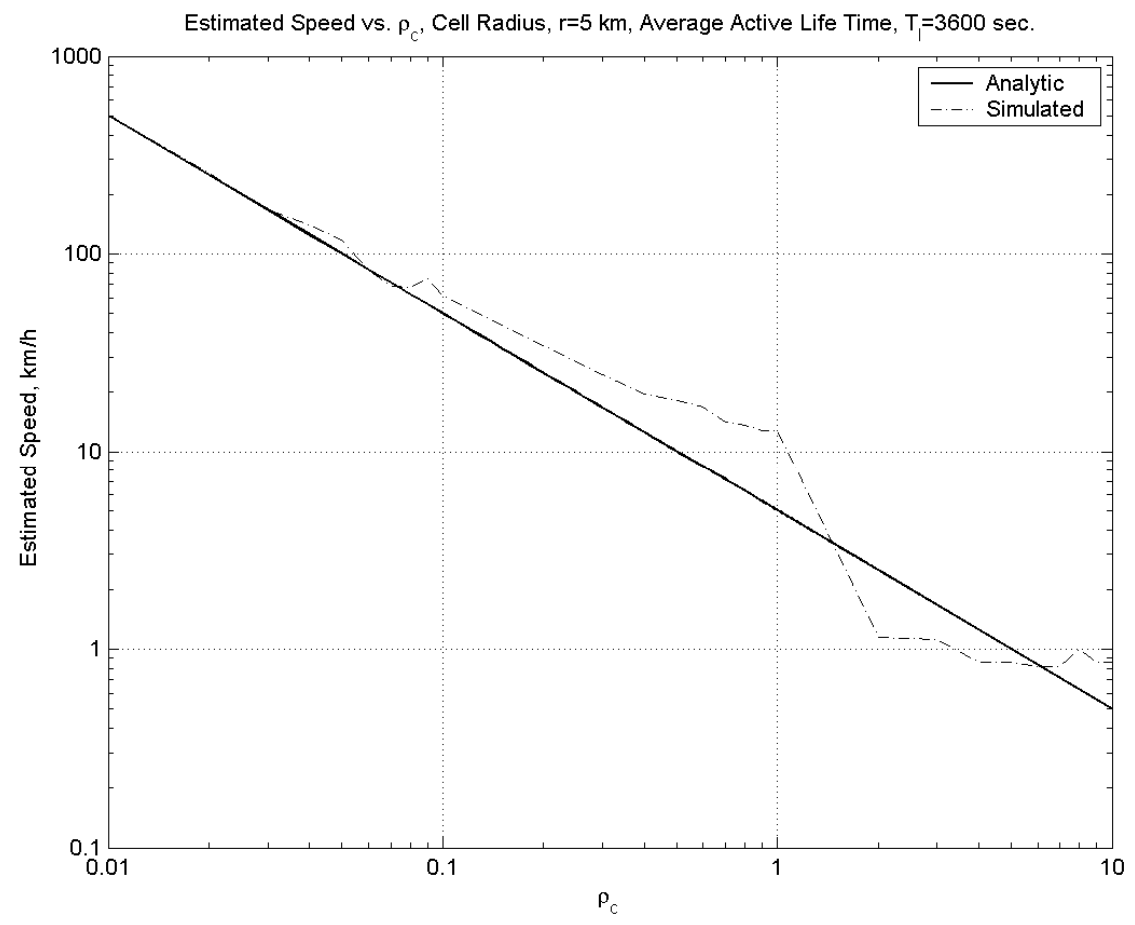

Figure 16. Relationship between the Life-to-Mobility Ratio and the estimated speed of mobile devices relative to RAN cell radius and average active life time.

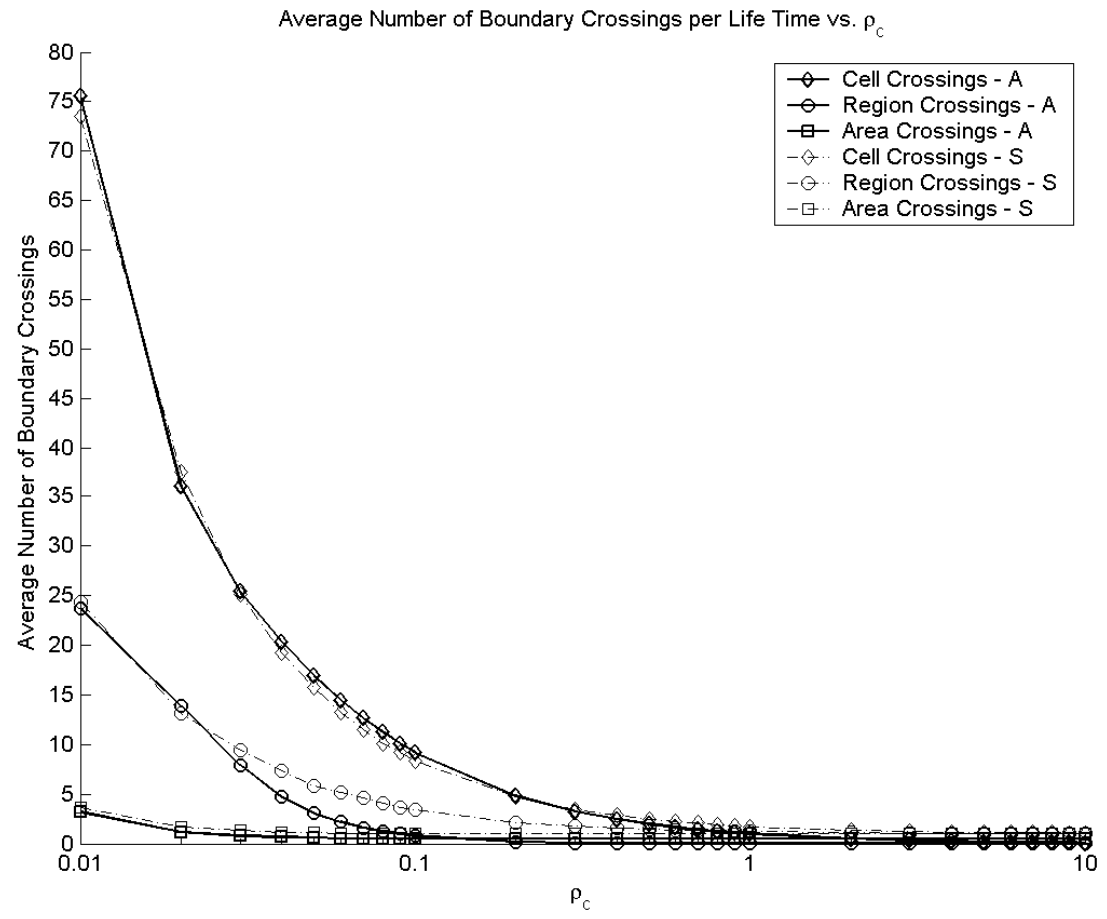

Figure 17. Average number of boundary crossings per active life time.

Figure 19b shows average excess network link utilization per an active session between two MNs. The excess link utilization is defined as the traffic that is carried by the additional network links required to serve the session due to the sub-optimal network path between the communicating MNs caused by the MIP triangular routing as compared to the optimal network path provided by H-MLBN. The average traffic rate of a session between the communication MNs is assumed to be $R = 64\ Kbps$. This excess traffic may also be called a network facing penalty due to triangular routing.

The user facing penalty for triangular routing may be defined as increased delay and jitter (delay variation), and increased packet loss probability caused by extra router hops and network links on the path between the communicating MNs.

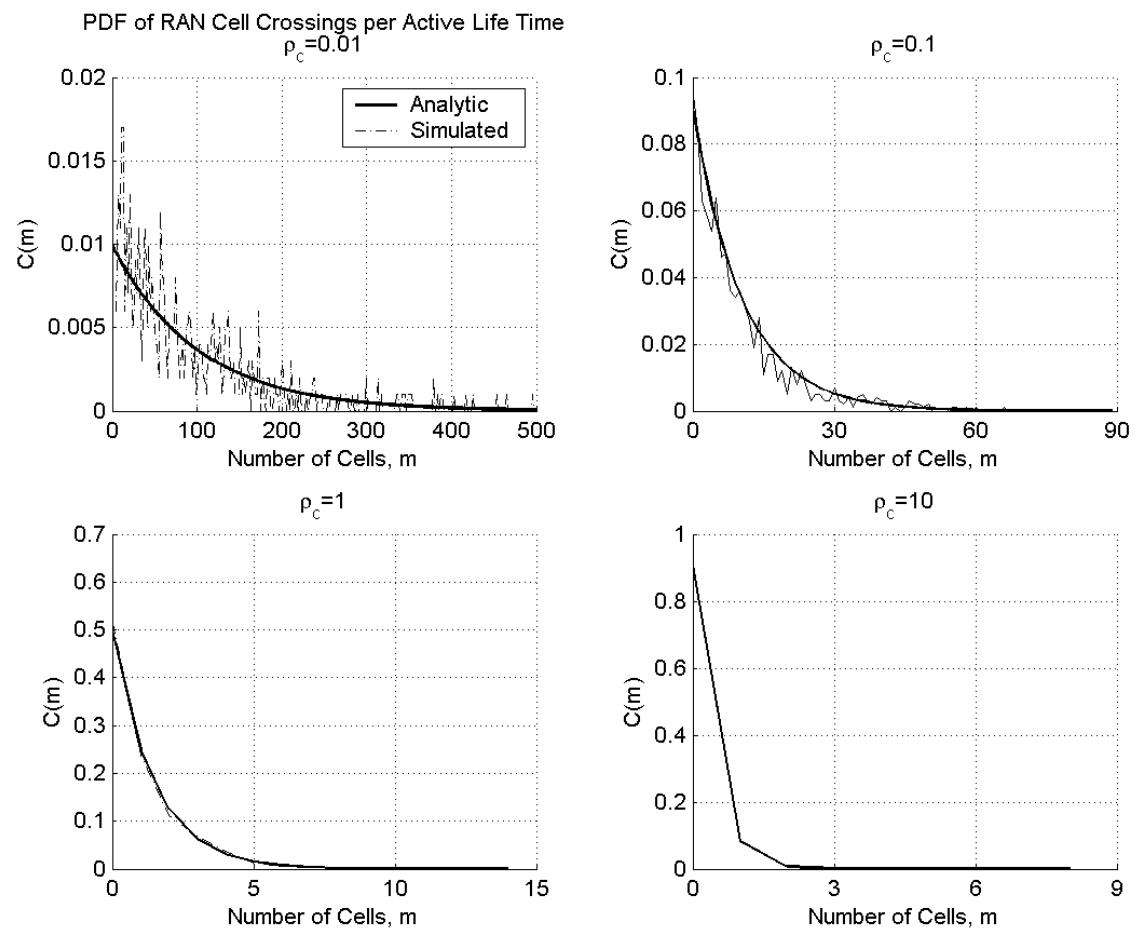

Figure 18a. Distributions of RAN cell boundary crossing probabilities for different mobility levels.

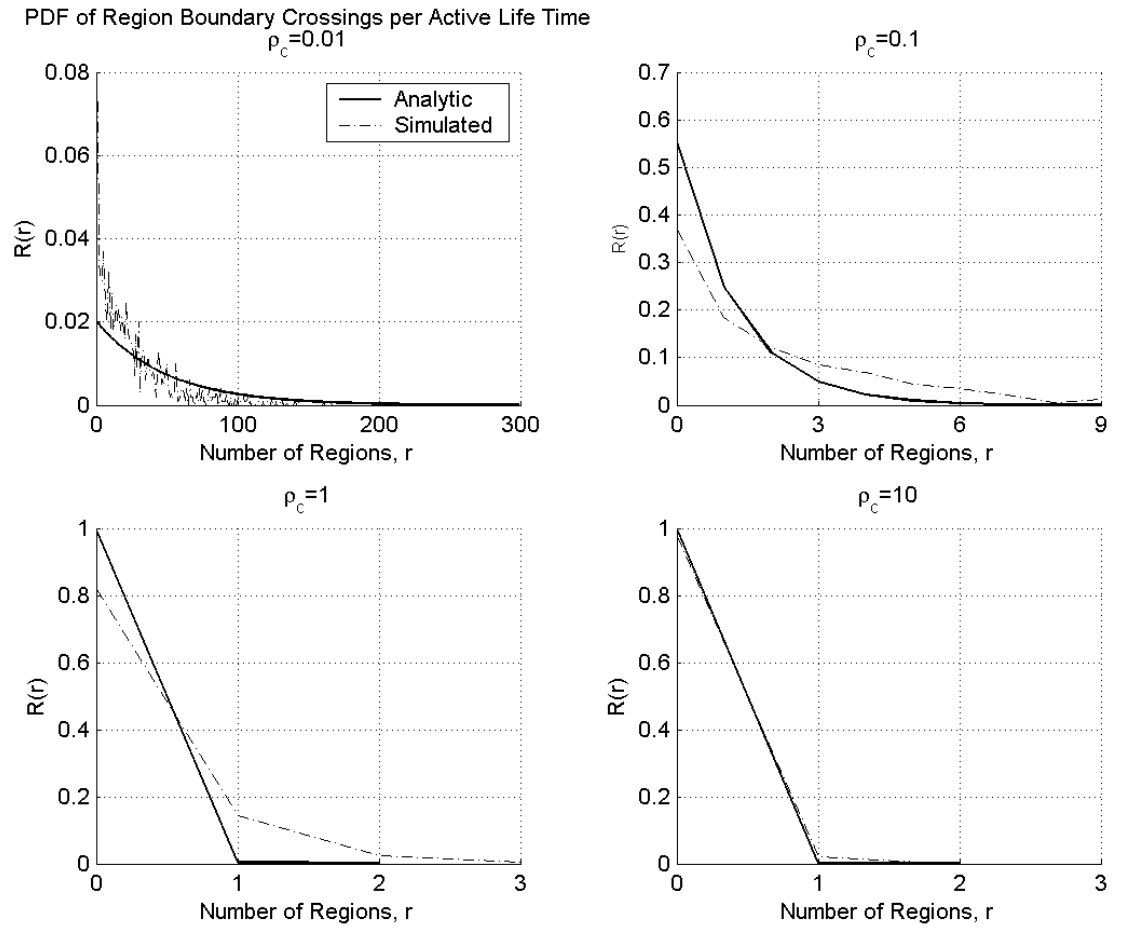

Figure 18b. Distributions of Mobility Region boundary crossing probabilities for different mobility levels.

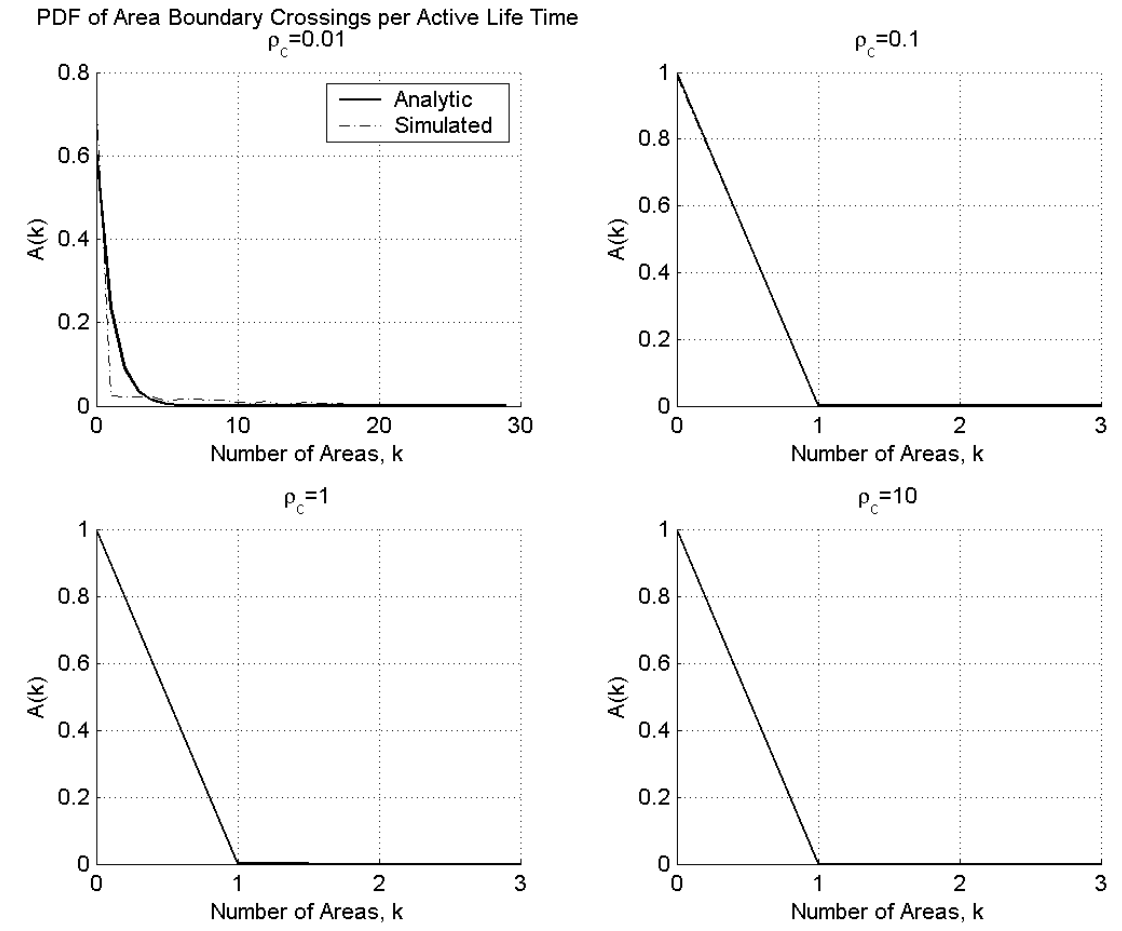

Figure 18c. Distributions of Mobility Area boundary crossing probabilities for different mobility levels.

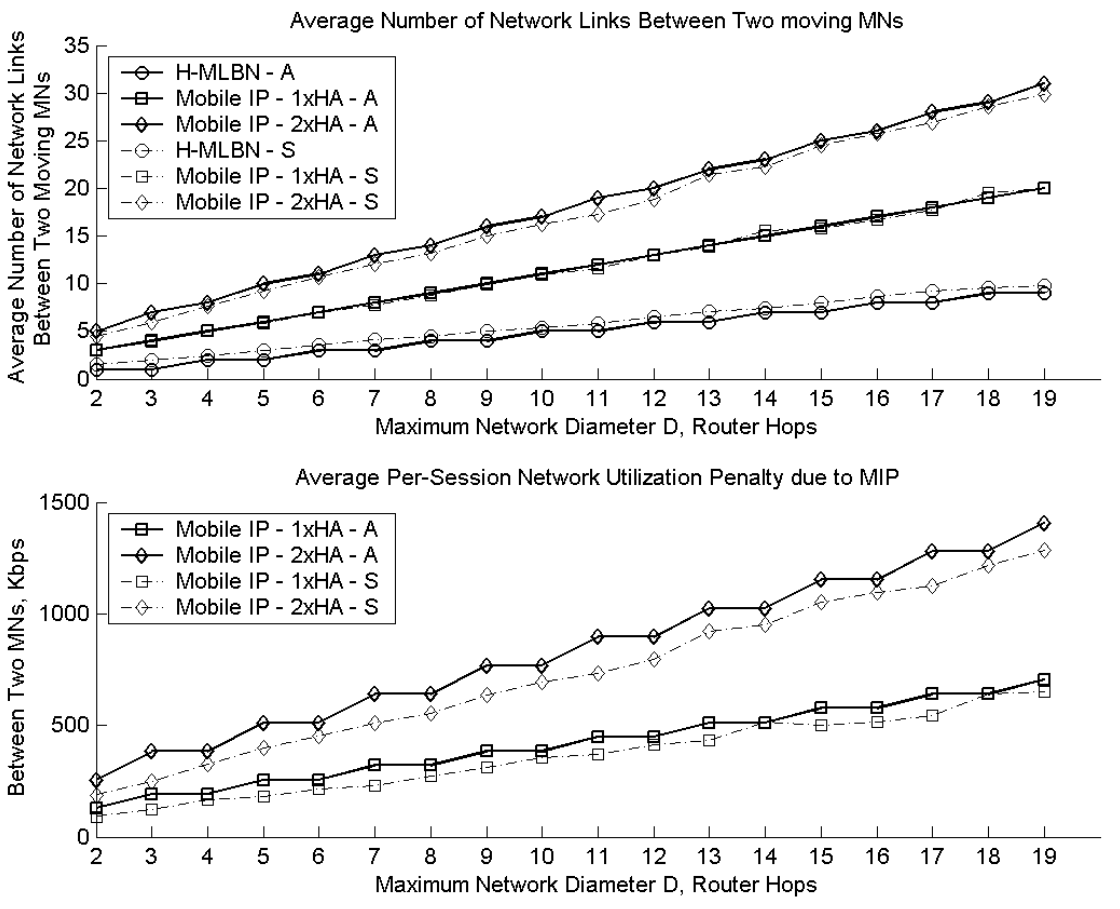

Figure 19. (a) Average network link count between communicating mobile nodes as a function of maximum network diameter in router hops., (b) average excess link utilization due to triangular routing.

As can be seen from Fig. 20, given the listed parameters, with a maximum network diameter of 10 hops the increase in one-way delay due to triangular routing for mobile-to-mobile communication is approximately between 30 and 60 milliseconds (depending on the number of serving HAs), along with the corresponding decrease in probability of successful packet delivery between 3 - 5%.

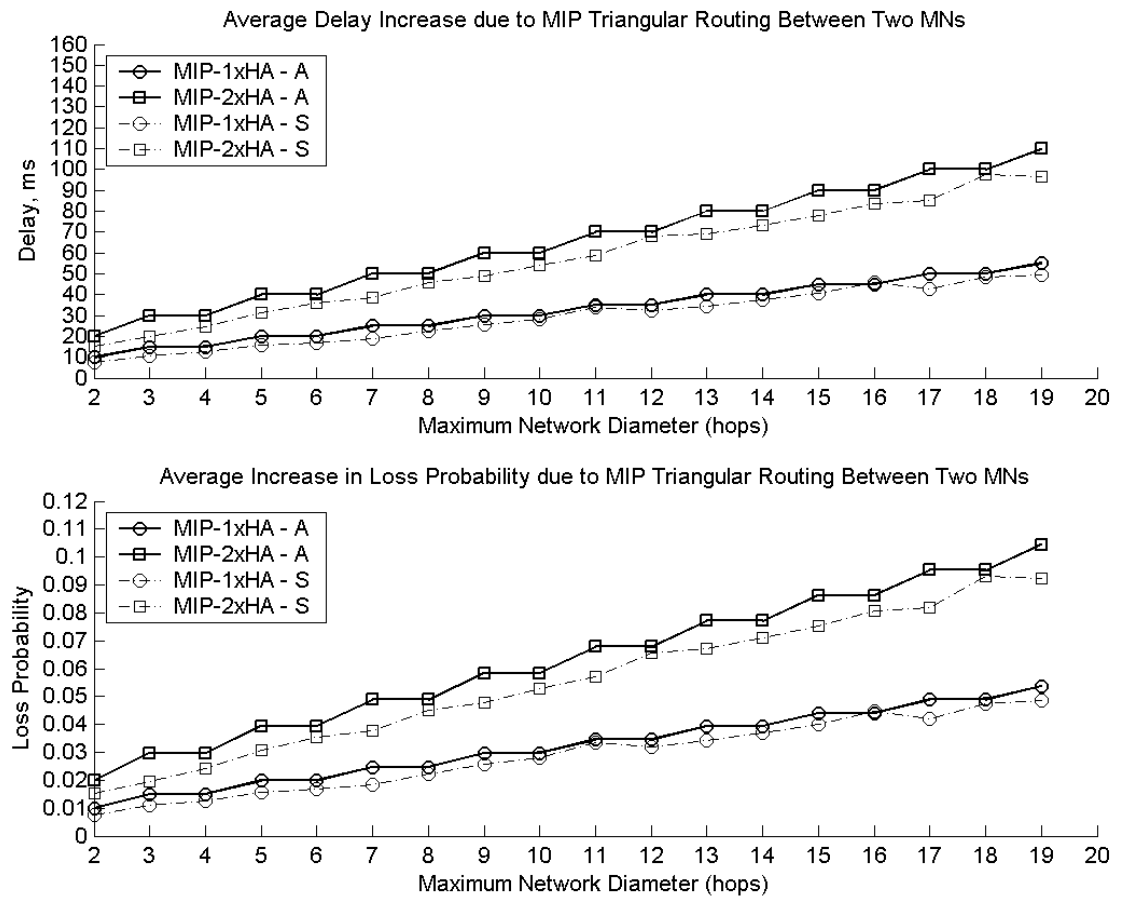

Figure 20. (a) Average increase in delay between communicating MNs due to triangular routing, (b) average increase in packet loss probability due to triangular routing.

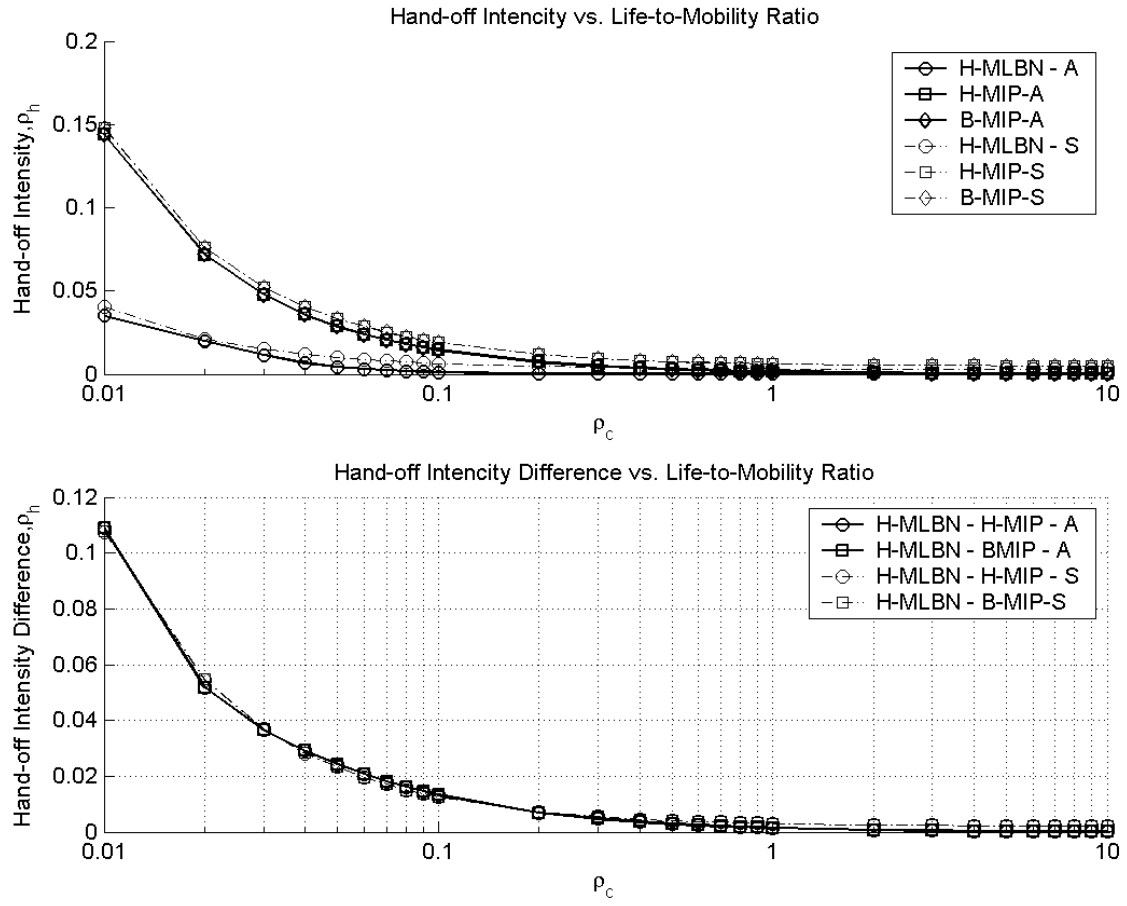

Figure 21. Hand-off Intensity $\rho_h$ vs. Life-to-Mobility Ratio $\rho_c$.

Fig. 21 provides comparison of the hand-off performance between H-MLBN and MIP based schemes. As can be seen, for the very high mobility level ($\rho_c = 0.01$) hand-off performance of both MIP based categories results in MNs spending on average approximately 15% of their active lives during the hand-off procedures, compared with 4% for H-MLBN. For the high mobility level ($\rho_c = 0.1$) the situation is similar: approximately 2% of active life is spent in hand-offs using MIP and 0.5% using H-MLBN.

Fig. 22 shows the update rates per type per mobility area. At the very high and high mobility levels the predominant type of updates are the local events caused by the cell boundary crossings internal to a MA. This figure also shows the MERS level (the combined local, intra-area and inter-area update rate corresponding to $\rho_c = 1$), which is equal to 100 updates per unit time per Mobility Area.

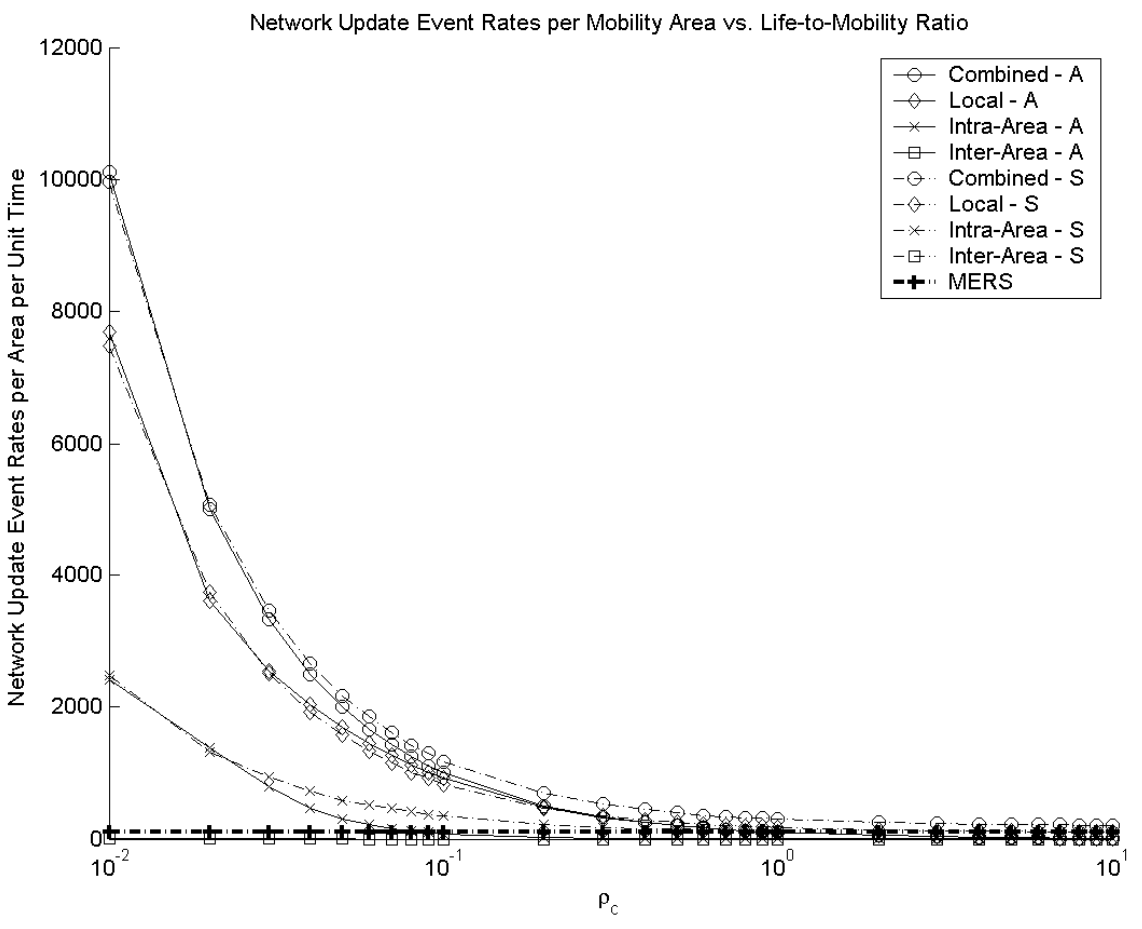

Figure 22. Network update event rates as a function of Life-to-Mobility Ratio.

Fig. 23 shows combined control plane message delivery cost as a function of Life-to-Mobility Ratio for the three mobility management categories: H-MLBN, H-MIP and B-MIP. Combined message delivery cost is defined as a sum of costs associated with the network update event types (local, intra-area and inter-area) measured in $\{hop \times message\ size \times event\ rate\}$, where the average hop count between network elements involved in the update process is determined by the network architecture, the average message size is determined by the protocol and the event rates depend on the mobility level. The costs and their differences are shown relative to the costs corresponding to MERS. Overall the H-MLBN scheme shows best cost performance across the mobility level range mostly due to the local update capability.

Similarly to Fig. 23, Fig. 24 shows the combined control plane message processing costs of the three mobility management solution categories as a function of the Life-to-Mobility Ratio. The message processing costs are measured in $\{cpu\ instructions \times event\ rate\}$, where CPU instructions depend on the cumulative number of instructions required to process the network update message by all network elements involved in the update and event rates depend on the mobility level. The combined processing cost is defined as a sum of the processing costs associated with the three network update types (local, intra-area and inter-area). As shown the H-MLBN solution results in lowest control plane processing costs across the mobility level range mostly due to the ability to perform low cost local update operations. Note that the processing costs for the H-MIP schemes are higher than for the B-MIP schemes at high mobility levels $(0.01 \leq \rho_c < 0.1)$ due to the increased number of regional and area hand-offs and the need to process updates at the FA, RFA and HA as compared to the FA and HA for B-MIP.

Finally, Fig. 25 shows the combined composite control plane messaging costs. The combined composite cost is a sum of the message delivery and the message processing costs. Overall H-MLBN produces the lowest control plane costs compared to MIP.

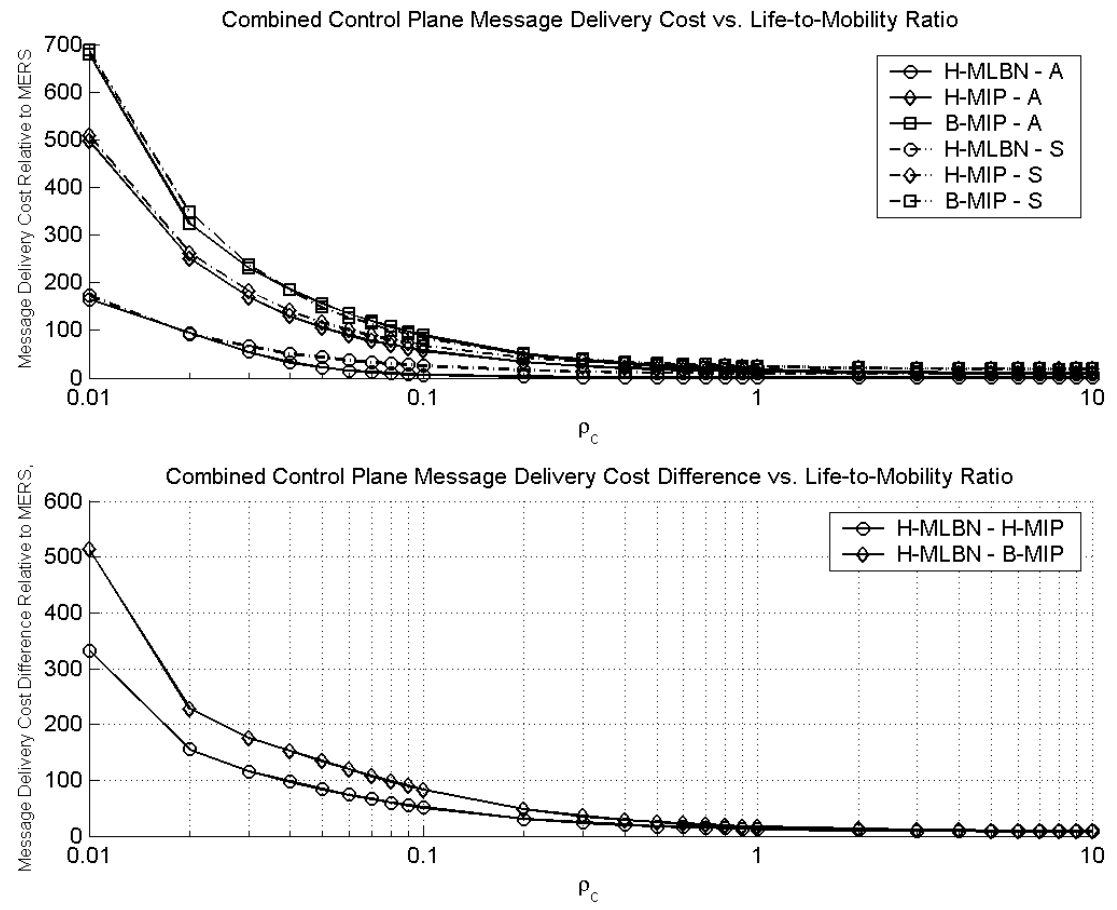

Figure 23. Combined control plane message delivery cost.

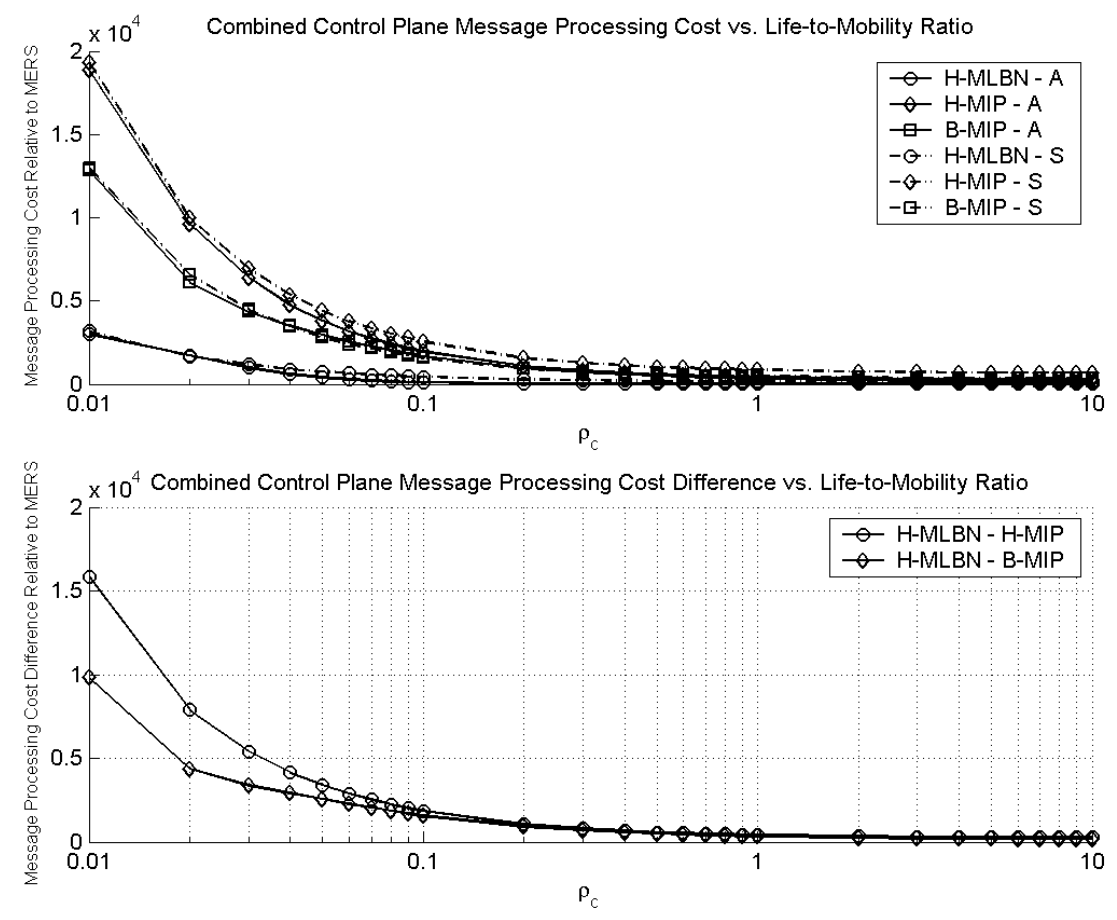

Figure 24. Combined control plane message processing cost.

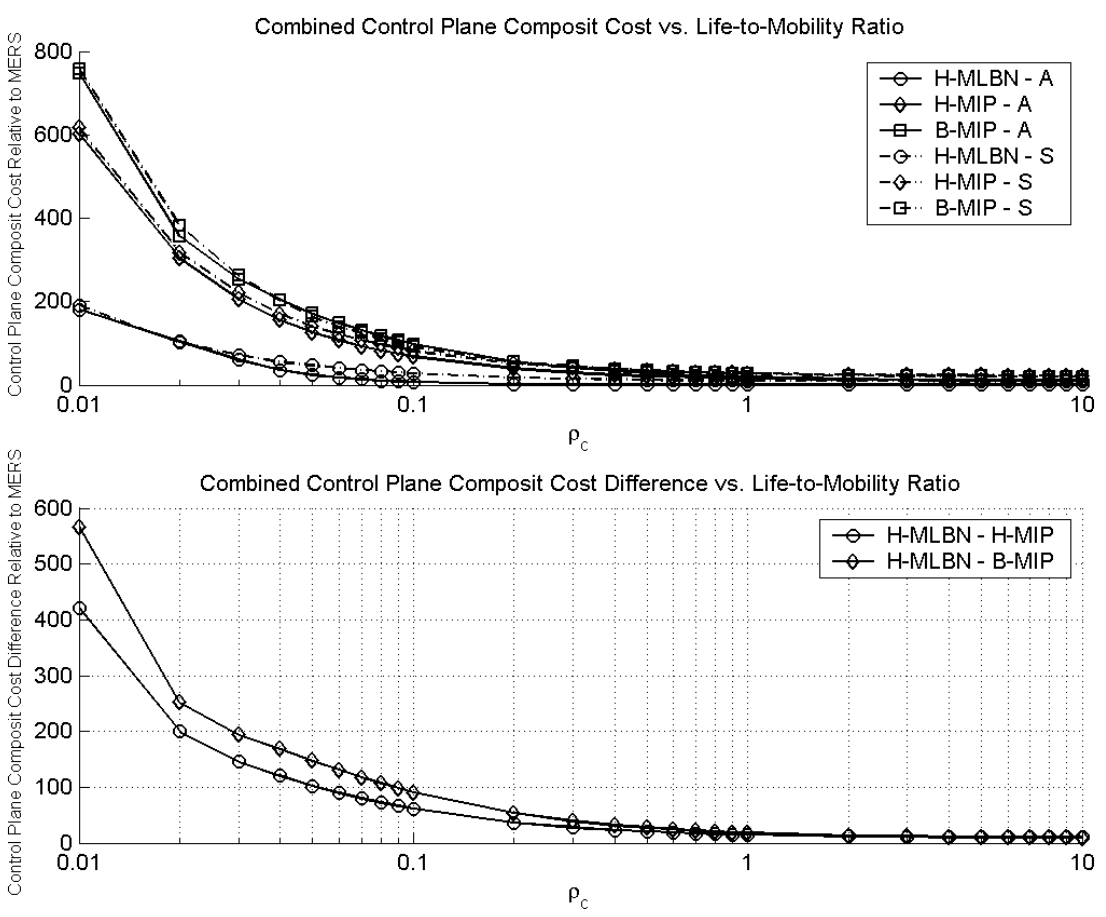

Figure 25. Combined composite control plane network update cost as a function of Life-to-Mobility Ratio.

## VII. Conclusions

The Hierarchical Mobility Label Based Network (H-MLBN) is a network layer mobility management system in which mobility control plane is fully integrated with MPLS forwarding plane resulting in optimal traffic delivery to mobile devices. H-MLBN provides support for macro- and micro-mobility for IPv4 and IPv6 mobile hosts and routers under a common MPLS-based control plane without a need for Mobile IP.

The use of MPLS-aware control plane and MPLS-based forwarding plane allows H-MLBN to provide optimally routed network path between the communicating mobile and fixed devices. This results in the ability to eliminate the user and network facing penalties associated with triangular routing and bidirectional tunneling commonly used in the Mobile IP based mobility management schemes. The presence of a Home Agent or multiple Home Agents in the session's traffic path results in additional network links and router hops that grow linearly with the network diameter.

The analytical and numerical results demonstrate that H-MLBN compares favorably with the competitive mobility management solutions especially in the area of optimal traffic delivery.

The hierarchical architecture and specific methods of providing support for micro-mobility in H-MLBN allow to control the cost of the network update messaging and shift the bulk of this cost onto the locally executed procedures performed by the edge nodes of the network. Specifically the local tracking of mobile devices by the H-MLBN LERs enables to use a low cost update procedure for the most frequent type of movement of mobile devices thus reducing the overall network update costs.

The analytical models derived in this paper for device movements as well as for the forwarding and control traffic delivery are independent from the mobility management solution and were used to provide comparative analysis which demonstrated significant performance improvements that can be achieved in H-MLBN over the Mobile IP based schemes both in the area of forwarding plane traffic delivery and the control plane messaging.

## References

[1]. O. Berzin, "Mobility Label Based Network: Mobility Support in Label Switched Networks with Multi-ProtocolBGP," Journal of Computer Networks (COMNET), Vol. 52, Issue 9, Jun. 2008, Page(s): 1732-1744.

[2]. O. Berzin, "Mobility Label Based Network: Hierarchical Mobility Management and Packet Forwarding Architecture," Journal of Computer Networks (COMNET), Vol. 53, Issue 12, Aug. 2009, Page(s): 2153-2181.

[3]. O. Berzin, A. Malis, "Mobility Support Using MPLS and MP-BGP Signaling", Internet Draft (work in progress), October 2007.

[4]. Perkins, "IP Mobility Support for IPv4", Request for Comment 3344, August 2002.

[5]. Rosen, Y. Rekhter, "BGP/MPLS VPNs", Request for Comment 4364, February 2006.

[6]. J. Schiller, "Mobile Communications", Second Edition, Addison-Wesley, 2003.

[7]. Y. Rekhter, E. Rosen, "Carrying Label Information in BGP-4," Request for Comment 3107, May 2001.

[8]. T. Bates et al., "Multiprotocol Extensions for BGP-4," Request for Comment 4760, January 2007.

[9]. A. G. Valko, "Cellular IP - A New Approach to Internet Host Mobility," ACM Computer Communication Review, January 1999.

[10]. E. Rosen et al., "Multiprotocol Label Switching Architecture", Request for Comment 3031, January 2001.

[11]. Johnson, D., Perkins, C., and J. Arkko, "Mobility Support in IPv6", Request for Comment 3775, June 2004.

[12]. F. M. Chiussi, D. A. Khotimsky, and S. Krishnan, "Mobility management in third-generation all-IP networks," IEEE Commun. Mag., pp. 124-135, Sep. 2002.

[13]. T. Yang, Y. Dong, Y. Zhang, and D. Makrakis, "Practical approaches for supporting micro mobility with MPLS," in IEEE ICT, 2002.

[14]. K. Xie, V.Wong, and V. Leung, "Support of micro-mobility in MPLS based wireless access networks," in Proc. IEEE WCNC,Mar. 2003, vol. 2, pp. 1242-1247.

[15]. V. Vassiliou, H. L. Owen, D. Barlow, J. Sokol, H.-P. Huth, and J. Grimminger, "M-MPLS: Micromobility-enabled multiprotocol label switching," in Proc. IEEE ICC, 2003, vol. 1, pp. 250-255.

[16]. R. Langar, S. Tohme, and N. Bouabdallah, "Mobility management support and performance analysis for wireless MPLS networks," ACMWiley Int. J. Netw. Manag., vol. 16, no. 4, pp. 279-294, Jul. 2006.

[17]. R. Langar, N. Bouabdallah, R. Boutaba, "A comprehensive analysis of mobility management in MPLS-based wireless access networks," IEEE/ACM Transactions on Networking, vol. 16, no 4, Aug. 2008.

[18]. C. Perkins, D. B. Johnson, "Route Optimization in Mobile IP," Internet draft, draft-ietf-mobileip-optim-11.txt, Sep. 2001.

[19]. K. Leung at al., "Network Mobility (NEMO) Extensions for Mobile IPv4", Request for Comment 5177, April 2008.

[20]. Devarapalli, V., Wakikawa, R., Petrescu, A., and P. Thubert, "Network Mobility (NEMO) Basic Support Protocol", Request for Comment 3963, January 2005.

[21]. E. Gustafson, A. Johnson, C. Perkins, "Mobile IP regional Registration," Internet draft, draft-ietf-mobileip-reg-tunnel-09.txt, June 2004.

[22]. Soliman et al., "Hierarchical MIPv6 Mobility Management," RFC 4140, Aug. 2005.

[23]. Chiussi, F.A.; Khotimsky, D.A.; Krishnan, S., "A network architecture for MPLS-based micro-mobility," Wireless Communications and Networking Conference, 2002. WCNC2002. 2002 IEEE, Volume: 2 Mar 2002, Page(s): 549-555 vol.2

[24]. K. El Malki, "Low-Latency Handoffs in Mobile IPv4," Request for Comment 4881, June 2007.

[25]. R. Koodli, "Fast Handovers for Mobile IPv6," Request for Comment 4068, July 2005.

[26]. S. Gundavelli et al., "Proxy Mobile IPv6", Request for Comment 5213, August 2008.

[27]. R. Ramaswamy, N. Weng, T. Wolf, "Considering processing cost in network simulations," ACM SIGCOM Workshops, Aug. 2003.

[28]. Y. Lin, "Reducing location update cost in a PCS network," IEEE/ACM Transactions on Networking, vol. 5, no. 1, Feb 1997.

[29]. R. Braden et al., "Resource ReSerVation Protocol (RSVP)," Request for Comment 2205, Sep. 1997.

[30]. L. Anderson et. al., "LDP Specification," Request for Comment 3036, Jan. 2001.